\documentclass[
reprint,
superscriptaddress,
amsmath,amssymb,
aps,
]{revtex4-2}

\usepackage{graphicx}
\usepackage{dcolumn}
\usepackage{bm}
\usepackage{color}
\usepackage{siunitx}
\usepackage{amsmath}
\usepackage{wasysym}
\usepackage{float}
\usepackage{array,url,kantlipsum}
\usepackage{verbatim}
\usepackage{xcolor}
\usepackage{CJK}
\usepackage{textcomp}
\newcolumntype{P}[1]{>{\centering\arraybackslash}p{#1}}
\usepackage{array}
\usepackage{upgreek}
\usepackage{scrextend}
\usepackage{hyperref}
\usepackage{multirow}

\hypersetup{
    colorlinks,
    citecolor=blue,
    filecolor=blue,
    linkcolor=blue,
    urlcolor=blue
}
\usepackage{braket}

\newcolumntype{x}[1]{>{\centering\arraybackslash\hspace{0pt}}p{#1}}

\begin{document}

\title{Excitation and probing of low-energy nuclear states at high-energy storage rings}

\author{Junlan Jin}
\affiliation{Department of Electrical and Computer Engineering, Princeton University, Princeton, New Jersey 08544, USA}
\affiliation{Department of Modern Physics, University of Science and Technology of China, Hefei 230026, China}

\author{Hendrik Bekker}
\affiliation{Johannes Gutenberg-Universit{\"a}t Mainz, 55128 Mainz, Germany}
 \affiliation{Helmholtz-Institut, GSI Helmholtzzentrum f{\"u}r Schwerionenforschung, 55128 Mainz, Germany}
 
\author{Tobias Kirschbaum}
\affiliation{Institute for Theoretical Physics und Astrophysics, University of W\"urzburg, 97074 Würzburg, Germany}

\author{Yuri A. Litvinov}
\affiliation{GSI Helmholtzzentrum f{\"u}r Schwerionenforschung, Planckstrasse 1, 64291 Darmstadt, Germany}

\author{Adriana P\'alffy}
\affiliation{Institute for Theoretical Physics und Astrophysics, University of W\"urzburg, 97074 Würzburg, Germany}

\author{Jonas Sommerfeldt}
\affiliation{Physikalisch-Technische Bundesanstalt, 38116 Braunschweig, Germany }
\affiliation{Technische Universit\"at Braunschweig, 38106 Braunschweig, Germany}

\author{Andrey Surzhykov}
\affiliation{Physikalisch-Technische Bundesanstalt, 38116 Braunschweig, Germany }
\affiliation{Technische Universit\"at Braunschweig, 38106 Braunschweig, Germany}

\author{Peter G. Thirolf}
\affiliation{Fakult\"at f\"ur Physik, Ludwig-Maximilians-Universit\"at M\"unchen, 85748 Garching, Germany}

\author{Dmitry Budker}
\affiliation{Johannes Gutenberg-Universit{\"a}t Mainz, 55128 Mainz, Germany}
\affiliation{Helmholtz-Institut, GSI Helmholtzzentrum f{\"u}r Schwerionenforschung, 55128 Mainz, Germany}
\affiliation{Department of Physics, University of California, Berkeley, California 94720, USA}

\date{\today}

\begin{abstract}

$^{229}$Th with a low-lying nuclear isomeric state is an essential candidate for a nuclear clock as well as many other applications. Laser excitation of the isomeric state has been a long-standing goal. With relativistic $^{229}$Th ions in storage rings, high-power lasers with wavelengths in the visible range or longer can be used to achieve high excitation rates of $^{229}$Th isomers.
This can be realized through direct resonant excitation, or excitation via an intermediate nuclear or electronic state, facilitated by the tunability of both the laser-beam and ion-bunch parameters.
Unique opportunities are offered by highly charged $^{229}$Th ions due to the
nuclear-state mixing.
The significantly reduced isomeric-state lifetime corresponds to a much higher excitation rate for direct resonant excitation.
Importantly, we propose electric dipole transitions changing both the electronic and nuclear states that are opened by the nuclear hyperfine mixing. We suggest using them for efficient isomer excitation in Li-like $^{229}$Th ions, via stimulated Raman adiabatic passage or single-laser excitation.
We also propose schemes for probing the isomers, utilizing nuclear radiative decay or laser spectroscopy on electronic transitions, through which the isomeric-state energy can be determined with an orders-of-magnitude higher precision than the current value.
The schemes proposed here for $^{229}$Th could also be adapted to low-energy nuclear states in other nuclei, such as $^{229}$Pa. 

\end{abstract}

\maketitle

\section{Introduction}

The first nuclear excited state in $^{229}$Th ($Z=90$), about 8\,eV above the ground state, has the lowest excited-state energy among all intrinsic nuclear excitations observed so far \cite{Seiferle2019energy,Sikorsky2020Energy,von2020ThEn}.
It also has a long radiative lifetime, on the order of $10^3 - 10^4$\,s \cite{Minkov_Palffy_PRL_2017}.
This unique nuclear isomeric state (denoted as $^{229m}$Th below) opens opportunities for building a high-precision nuclear clock, testing fundamental physics, as well as for other interdisciplinary research such as investigating the interplay between the nuclear and electronic degrees of freedom of the atom; see, for example, Refs.\,\cite{Beeks2021Th,Peik2021} and references therein.

There has been significant effort in pursuing excitation and detection of $^{229m}$Th \cite{Jeet2015DirectExc,Yamaguchi2015DirectExc,Stellmer2018OpticalExc,Wense2017DirectExc,von2020frequency-comb}.
For instance, excitation of $^{229}$Th ions doped in a crystal host using vacuum ultraviolet photons was experimentally attempted \cite{Jeet2015DirectExc}, but population of the isomeric state was not observed. There have also been proposals of using the internal conversion (IC) in neutral thorium atoms \cite{Wense2017DirectExc}, and using the electron-bridge process in low-charge \cite{Porsev2010EB,Bilous2018EB} or highly-charged  $^{229}$Th ions ($^{229}$Th$^{35+}$ in Ref.\,\cite{Bilous2020EB_35+}) for investigating isomer excitation or decay.
$^{229m}$Th can also be produced through excitation of the second nuclear exited state at 29\,keV.
X-ray pumping of the isomeric state using this approach was demonstrated in Ref.\,\cite{Masuda2019ThXray}.
Recently, the radiative decay of $^{229m}$Th was observed at the ISOLDE facility at CERN \cite{Kraemer2022_obs_rad_decay}. However, laser excitation of $^{229m}$Th has not been realized yet, and the isomeric-state energy has not been determined precisely (only with a relative uncertainty of $2\%$ in Refs.\,\cite{Seiferle2019energy,Sikorsky2020Energy} \footnote{0.3\% was recently reported in Ref.\,\cite{Kraemer2022_obs_rad_decay}}).

Most studies on $^{229m}$Th have dealt with nonrelativistic atoms or ions in low charge states, posing high demand on the light source required for excitation.
This manuscript surveys the possibilities of laser excitation and probing of $^{229m}$Th using relativistic highly charged ions (HCI) in high-energy storage rings, leveraging the relativistic Doppler effect and the significant nuclear-state mixing in highly charged $^{229}$Th ions.

The energy of relativistic ions in storage rings can be flexibly adjusted, complementary to tuning the laser photon energy. With the Lorentz boost of photon energies in the ion frame, high-power lasers with wavelengths in the visible range or longer combined with a Fabry–P\'erot cavity can be used to obtain high isomer-excitation rates. 
Remarkably, even higher excitation rates and more unique opportunities become available when using H- or Li-like thorium ions.
The tightly bound unpaired electron in those ions produces a strong magnetic field at the nucleus, mixing hyperfine levels corresponding to different nuclear states, known as the nuclear hyperfine mixing (NHM) \cite{Karpeshin1998NHM,Tkalya2016_NHM,Shabaev2021NHM}.
The NHM effect could reduce the isomeric-state lifetime
by orders of magnitude, which leads to significantly higher excitation rates in the direct resonant excitation and could enable implementing a $\pi-$pulse.
Furthermore, as a key point of this work, we propose electric-dipole ($E1$) forbidden transitions in Li-like $^{229}$Th ions that are opened by the NHM (Sec.\,\ref{Subsec:exc_271eV}). We present isomer excitation schemes using those transitions, including both stimulated Raman adiabatic passage (STIRAP) and single laser excitation.
An encouraging aspect of the proposal is that it employs only the technologies already available today.
We also discuss direct resonant excitation in bare nuclei and excitation via the second nuclear excited state. Candidate storage rings for implementing those schemes can be, for instance, those at GSI/FAIR
\cite{Litvinov-2013,Shevelko2018IonLifetimeHESR,Sanchez2020HESR,Spiller2020SIS100},
HIAF~\cite{Yang-2013},
and the proposed Gamma Factory (GF) at CERN \cite{Krasny:2015ffb,Krasny2019PoP,jaeckel2020quest,Budker2020_AdP_GF,Budker2021GF_nucl}.

Among the discussed excitation schemes, exploiting the NHM effect could be most favorable and efficient for experimental implementation (Sec.\,\ref{subsubsec:res_H-like} and \ref{Subsec:exc_271eV}).
Schemes utilizing repeated pumping can significantly relax the demand on the laser-pulse intensity by using a longer pulse sequence (Sec.\,\ref{subsubsec:reso_multi_pulse} and \ref{subsubsec:inco_exc_29keV}).
STIRAP excitation via the second nuclear excited state would require further development of the laser or storage-ring technologies (Sec.\,\ref{subsubsec:STIRAP_29keV}). 
Schemes using an intermediate state could also be realized by using x-ray pulses combined with static or moderately accelerated $^{229}$Th ions (Sec.\,\ref{subsec:exc_x_ray}). 

We also propose schemes for detecting the isomers in high-energy storage rings, including one dedicated to Li-like $^{229}$Th, where laser spectroscopy on electronic transitions allows for distinguishing the nuclear ground and isomeric states (Sec.\,\ref{Subsec:elec_tran}).
We also suggest determining the isomeric-state energy using the recent precision measurement of the energy of the second nuclear excited state (Sec.\,\ref{Subsec:Fur_exc}).
According to our estimate, the energy of the isomeric state can be measured with a precision better than $10^{-4}$ or even down to
below $10^{-6}$, which is orders of magnitude improvement of the present value.
This would pave the way for further improvement in determining the isomeric-state energy and bring a precise nuclear clock and fundamental-physics experiments with this system closer to reality.

This manuscript is structured as follows.
In Sec.\,\ref{Sec:ion_source}, we introduce possible ways to obtain thorium ions stored in a high-energy storage ring and ion-bunch parameters for various operational or planned facilities.
We discuss schemes for isomer excitation in Sec.\,\ref{Sec:iso_exc}, sorted by the energy of corresponding transitions. Methods for probing produced isomers are presented in Sec.\,\ref{Sec:Detect}. We discuss searching for the resonant excitation in Sec.\,\ref{Sec:search_res}. Extending the schemes to other low-energy nuclear states is discussed in Sec.\,\ref{Sec:other_low_state}.

\section{Preparation of stored and cooled intense beams of Thorium-229}
\label{Sec:ion_source}

There are several technological prerequisites for the studies proposed to become possible.

One of the major challenges is to obtain HCI stored in  accelerators. An efficient method for producing beams of heavy HCI is to start with low-charge ions from an ion source, and then accelerate them to high energies before focusing onto a stripper target to remove more of the bound electrons; see, for example, Refs.\,\cite{Atanasov-2015, Steck2020HeavyIonRing,Kroger2021PSI_source}.
The relatively long half-life of about 7880\,y makes it possible in principle to prepare a $^{229}$Th beam directly from an ion source, like it is done for instance for beams of $^7$Be\,\cite{Limata-2008}.
In such case, $^{229}$Th nuclei need to be produced via irradiation of uranium or thorium targets and chemically purified from inevitable contaminants.
Still, the electrodes will be highly active and the overall process is technologically challenging.

Alternatively, a projectile-fragmentation nuclear reaction can be employed for in-flight production of $^{229}$Th beams.
Here, either $^{238}$U or $^{232}$Th primary beams can be accelerated and fragmented in a target, thereby creating $^{229}$Th ions of interest. The kinetic energy of the beam together with the target material and its thickness can be chosen such that the Th ions emerge from
the target in the desired atomic charge state\,\cite{Litvinov-Bosch-2011}.
Assuming, for example, a $^{238}$U ($^{232}$Th) beam at 600\,A\,MeV with a realistic intensity of 10$^{10}$ particles/s impinging on a 1\,g/cm$^2$ $^9$Be production target, we obtain after the target $3\cdot10^5$ ($3\cdot10^6$)  $^{229}$Th ions per second \cite{LISE}. These rates can easily be scaled if higher primary beam intensities are available.
Accumulation schemes would be needed to achieve the required $^{229}$Th intensities.

Another issue is related to the transport of laser beams into the ultra-high vacuum environment of the ring. 
Various solutions have been realized at several facilities, see, for example, Ref.\,\cite{Nortershauser-2015}. A high-photon-flux extreme ultraviolet source based on high harmonic generation was recently commissioned at the CRYRING\,\cite{Hilbert-2019, Hilbert-2020}. For this purpose viewports are placed at dedicated locations to enable the spacial overlap of the stored beam with the injected laser photons. 
Since the photon beam cannot be bent, a special design of bending dipole magnets is required, which might be a challenging task for already existing machines and/or if superconducting magnets are considered\,\cite{Winters-2015a}.

Presently, stored and cooled beams of highly-charged stable or radioactive ions are routinely available at GSI \cite{Franzke-2008}. The experimental storage ring ESR \cite{Franzke-1987} offers beams at energies of about 4-400\,A\,MeV for a variety of precision studies\,\cite{Litvinov-2013,Bosch-2013}. Relevant here are the possibilities to collide the stored beams with photon\,\cite{Ullmann-2017}, electron\,\cite{Brandau-2009}, or atomic\,\cite{Glorius-2019} targets. 
The production and in-ring purification of stored beams of secondary $^{234}$Pa and $^{237}$U ions was demonstrated in Refs.\,\cite{Brandau-2009,Brandau-2010}. Following these experiments, a pure beam of few-electron $^{229}$Th ions was prepared in the ESR in 2022\,\cite{Brandau-PC}. Also an efficient accumulation scheme is established at the ESR\,\cite{Steck2020HeavyIonRing}. At lower energies, the CRYRING\,\cite{Lestinsky-2016} facility can be employed. We also note a proposal to study $^{229m}$Th in the experimental cooler-storage ring CSRe at IMP in China\,\cite{Ma-2015}.
Importantly, the ESR, CRYRING and CSRe are specially designed for laser-spectroscopy studies.

As of today, no facility exists to provide stored and cooled beams of $^{229}$Th at much higher energies. Beams with energies of up to 5\,A\,GeV will be available at the FAIR facility in Darmstadt. The design parameters are to provide accelerated $^{238}$U$^{28+}$ beams with an intensity of 10$^{13}$ particles/s \cite{Durante-2019}. Technically, acceleration of $^{232}$Th beams is also possible. The secondary $^{229}$Th ions produced from projectile fragmentation will be separated from contaminants via the high-acceptance in-flight fragment separator Super-FRS\,\cite{Geissel-2003s} and injected into the collector ring CR\,\cite{cr}. 
The CR is designed for fast stochastic cooling of secondary beams.
Afterwards, the beam will be transferred to and accumulated in the high-energy storage ring HESR\,\cite{hesr}  for precision experiments.
An additional storage ring, RESR,
between the CR and HESR, is foreseen at a later stage of FAIR to achieve a more efficient beam accumulation. 
Two $0^\circ$ laser-ion intersection regions are being prepared in the HESR\,\cite{Sanchez2020HESR, Stohlker-2014,Stohlker-2015,Stohlker-2015b}. A similar scheme will be possible to realize at the HIAF facility under construction in China\,\cite{Yang-2013}.

At CERN, the in-flight production of $^{229}$Th beam can be envisioned between the Proton Synchrotron (PS) and the Super Proton Synchrotron (SPS). Also the isotope separation on-line (ISOL)  production method can be employed as utilized at the ISOLDE facility\,\cite{Catherall2017_ISOLDE, Borge_2017}. Here, a post-accelerated to 10\,A\,MeV $^{229}$Th beam will need to be transported from the ISOLDE hall to the CERN accelerator chain. The advantage of this approach would be the much larger $^{229}$Th production yields owing to several hundred times thicker production targets. The proposal for the Gamma Factory\,\cite{Budker2021GF_nucl} also contains a dedicated storage ring for storing beams of radioactive nuclei from ISOLDE. In such configuration, the beams of $^{229}$Th ions will be stored at about 10\,A\,MeV energy\,\cite{Grieser-2012} and irradiated with high-energy photons from the GF.

Parameters of some synchrotron facilities and ion storage rings are listed in Table\,\ref{tab:ion-beam_facilities} sorted by the relativistic Lorentz factor $\gamma=1/(1-\beta^2)^{1/2}$ ($\beta=v/c$ with $v$ and $c$ being the speed of ions and light, respectively). Technical solutions for production and cooling of highly-charged $^{229}$Th beams and/or coupling of laser beams are in place for the ESR and are planned for HESR and BRING. This would also need to be done for the other machines.
It is not our aim in the present work to investigate possible upgrades of these facilities, but to show that the challenges can be met at the present level of technology.
The parameters in Table\,\ref{tab:ion-beam_facilities} assume a stored beam of $^{229}$Th$^{q+}$ ions ($q\approx90$), though such beams cannot presently be provided at some of the listed facilities.

\begin{table*}[htpb]
    \centering
    \begin{tabular*}{\linewidth}{@{\extracolsep{\fill}} llccc clcl}
    \hline 
    \hline
    Facility    &Lab.   &$\gamma_\textrm{max}$  &$\Delta\gamma/\gamma$   &$N_\textrm{ion}$/bunch  &1-$\sigma$ radius  &Circumference, $f_b$   &$N_b$ \\
    \hline \\[-0.2cm]
    LHC\,\cite{Budker2021GF_nucl}
    &CERN      &2950 &$\sim10^{-4}$--($10^{-6}$)  &$10^8$   &16\,$\mu$m  &26.7\,km, 11.2\,kHz  &592--1232\\
    SPS\,\cite{Krasny2019PoP}\,\footnote{The parameters are from Ref.\,\cite{Krasny2019PoP} for Li-like Pb ions. Beam parameters for Li-like Th ions should be similar.}      &CERN       &109   &$\sim10^{-4}$--($10^{-6}$)  &$10^8$   &0.5\,mm  &6911\,m, 43.4\,kHz  &$\leq$36\\
    RHIC\,\cite{Workman2022_Particle_Data,RHIC-2003}      &BNL     &100  &$7.5\cdot10^{-4}$  &$4.5\cdot10^{10}$   &100\,$\mu$m  &3834\,m, 78.2\,kHz  &106\\
    SIS100\,\cite{Winters-2015a,Spiller2020SIS100}    &FAIR\,\footnote{\label{construct}Under construction.}   &100    &10$^{-3}$--(10$^{-7}$)  &10$^{13}$  &  $\sim$mm\,\footnote{\label{beamsize}A rough estimate. The beam size is determined by the intra-beam repulsion and counteracting cooling force\,\cite{Steck2020HeavyIonRing}. Beam envelope varies along the machine lattice dependent on the Twiss functions. No dedicated focus is presently considered in the optics design of this facility. }
    &1084\,m, 280\,kHz  & 1\,\footnote{\label{exciter}Additional bunching can be achieved by a dedicated exciter creating up to several tens of bunches\,\cite{Winters-2015a}, with the total number of stored ions remaining the same.}\,\footnote{\label{short}Bunches with a duration of several ns can be generated using a bunch-compression scheme.}\\
    Main Ring\,\cite{JPARC,JPARC3,JPARC4}     & J-PARC-HI\,\footref{construct}      & 22  & $\sim10^{-3}$\,\footnote{\label{guess}Rough estimate based on a typical synchrotron performance.} & $10^{11}$  & $\sim$mm\,\footref{beamsize} &1567.5\,m, 191\,kHz & 8\\
    PS\,\cite{PS-2011}    &CERN      & 11.4 & $\sim10^{-3}$\,\footref{guess} & $10^{14}$  & $\sim$mm\,\footref{beamsize} &628\,m, 478\,kHz  & $\leq4$\\
    HESR\,\cite{Kovalenko2015HESR,Shevelko2018IonLifetimeHESR,Sanchez2020HESR}     &FAIR\,\footref{construct}    &6    &$\sim10^{-5}$  &$10^8$  &$\sim2.1$mm\,\footref{beamsize}   &575\,m, 522\,kHz  &1\,\footnote{\label{coasting}Coasting or unbunched ion beams, i.e., the ion beam filling the whole storage ring almost uniformly.}\\
    BRING1\,\cite{Yang-2013} & HIAF\,\footref{construct} & 4.4 & $\sim10^{-3}$\,\footref{guess} & $2\cdot10^{11}$ & $\sim$mm\,\footref{beamsize} & 600\,m, 490\,kHz & 1\,\footref{exciter} \\
    SRING\,\cite{Yang-2013} & HIAF\,\footref{construct} & 2.1 $(\beta=0.879)$ & $\sim10^{-5}$ & 10$^8$ & $\sim$ mm\,\footref{beamsize} & 273\,m, 966\,kHz & 1-2\,\footref{coasting}\,\footref{exciter} \\
    ESR\,\cite{Franzke-1987} &GSI &1.5 $(\beta=0.745)$ &$\sim10^{-5}$ &$10^8$  &$\sim$mm\,\footref{beamsize} &108\,m, 2.07\,MHz &1-2\,\footref{coasting}\,\footref{exciter}\\
    \hline
    \hline
    \end{tabular*}
    \caption{Relativistic HCI beams expected at different accelerators. 
    Here, $\gamma_\textrm{max}$ is the maximal relativistic Lorentz factor of
    highly charged $^{229}$Th$^{q+}$ ions ($q\approx90$ in this work). 
    The Lorentz boost of laser photon energies is $(1+\beta)\cdot\gamma$ in the ion frame.
    $\Delta\gamma/\gamma$ is roughly the FWHM relative range of $\gamma$, due to the ion energy spread.
    The lower bound of $\Delta\gamma/\gamma$ may be reached by laser cooling, which might require reducing the intensity of the ion bunch to alleviate intrabeam scattering \cite{Eidam2018LaserCooling}.
    The number of particles $N_\textrm{ion}$ per bunch relates to expected maximum particle intensities that can be stored in the machine, irrespective to the actual production method of $^{229}$Th beams.
    }
    \label{tab:ion-beam_facilities}
\end{table*}

\section{Excitation of the thorium isomer}
\label{Sec:iso_exc}

\begin{figure}[!htpb]\centering
    \includegraphics[width=\linewidth]{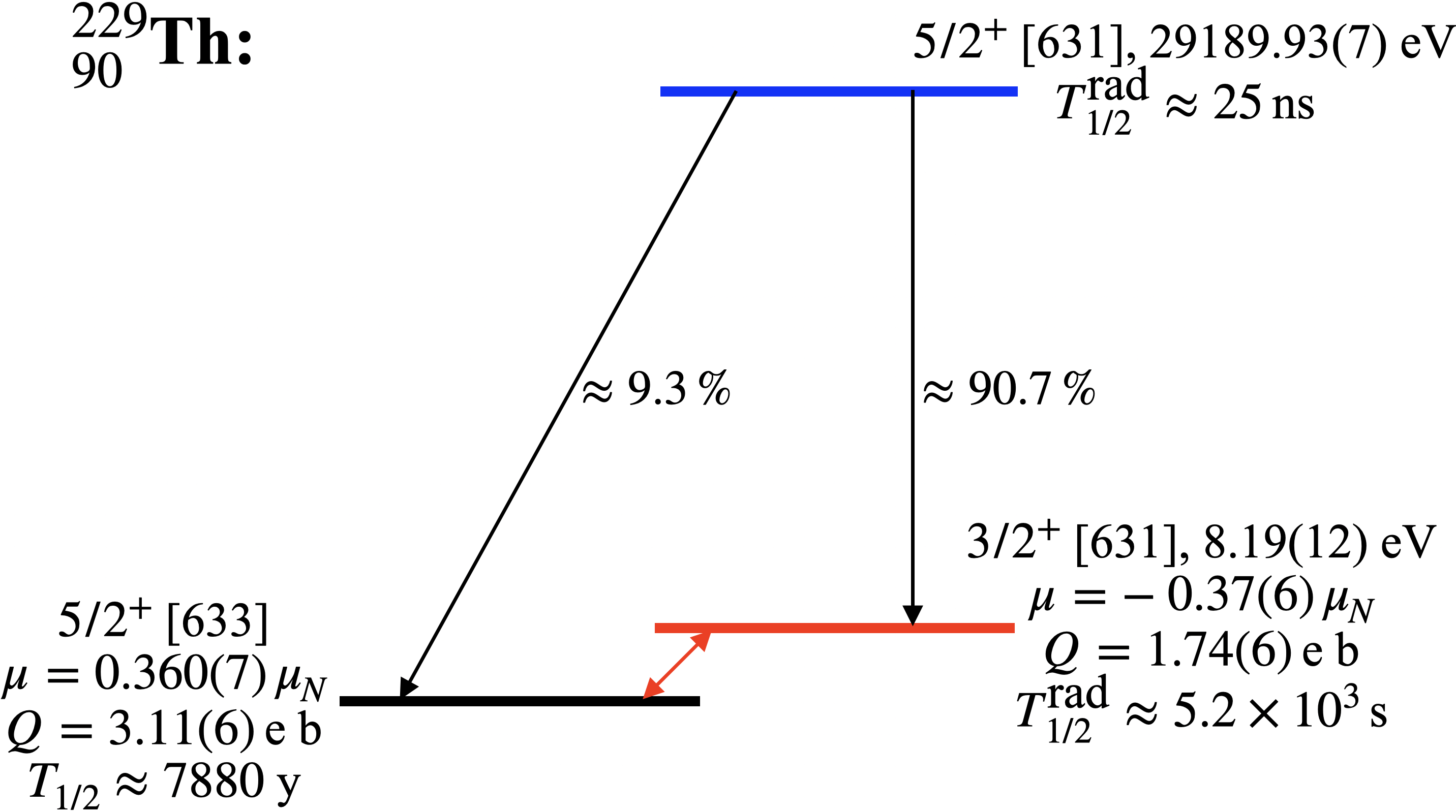}
    \caption{Low-lying nuclear levels of $^{229}$Th. These levels belong to two rotational bands (indicated by the horizontal offset) labelled by spin, parity, Nilsson quantum numbers, energy and (radiative) half-life $T_{1/2}^{(\textrm{rad})}$. $T_{1/2}^{\textrm{rad}}$ of the isomeric state is derived from Ref.\,\cite{Minkov_Palffy_PRL_2017}.
    $T_{1/2}^{\textrm{rad}}$ of the second excited state is obtained using experimental results in Refs.\,\cite{Masuda2019ThXray,Sikorsky2020Energy}. 9.3\% and 90.7\% are the radiative branching ratios.
    The nuclear magnetic dipole moment $\mu$ and electric quadrupole moment $Q$ are presented for the ground and isomeric states \cite{Campbell2011Th,Safronova2013MD_EQ,Thielking2018laser}.
}
    \label{fig:229Th}
\end{figure}

We begin with introducing the low-lying nuclear energy levels of $^{229}$Th related to this work (see Fig.\,\ref{fig:229Th}).
The isomeric state has an energy of $E_{\textrm{iso}}=8.19(12)$\,eV \cite{Peik2021}, an average of two experimental values from Refs.\,\cite{Seiferle2019energy,Sikorsky2020Energy}.
The radiative half-life  of the isomeric state in bare nuclei is $T_{1/2}^{\textrm{rad}}\approx5.2\times10^3\,$s, derived theoretically in Ref.\,\cite{Minkov_Palffy_PRL_2017} using a reduced transition probability $B(M1)\approx0.0076\,$W.u. (Weisskopf units).
The second nuclear excited state, with an energy of $E_{\textrm{2nd}}=29,189.93(7)$ \cite{Masuda2019ThXray} and a radiative half-life of about $25\,$ns, can likewise only decay via radiative transitions, while the internal conversion channel is energetically forbidden in highly charged thorium ions discussed in this work.
The cross-band transition from the second excited state to the ground state has a radiative width of $\Gamma_{\gamma}^{\textrm{cr}}=1.70(40)\,$neV, determined experimentally in Ref.\,\cite{Masuda2019ThXray}. The branching ratio of this transition is BR$^{\textrm{cr}}_\gamma=9.3(6)\%$ \cite{Sikorsky2020Energy}. We can further derive the branching ratio, BR$^{\textrm{in}}_\gamma=90.7(6)\%$, and radiative width, $\Gamma_{\gamma}^{\textrm{in}}=16.6(41)\,$neV, for the in-band transition, from the second excited state to the isomeric state (see also Ref.\,\cite{Kirschbaum2022}).
The nuclear magnetic dipole moment $\mu$ and electric quadrupole moment $Q$ of the ground and isomeric states shown in Fig.\,\ref{fig:229Th} are from Refs.\,\cite{Campbell2011Th,Safronova2013MD_EQ,Thielking2018laser}.

\begin{figure*}[!htpb]\centering
    \includegraphics[width=0.95\textwidth]{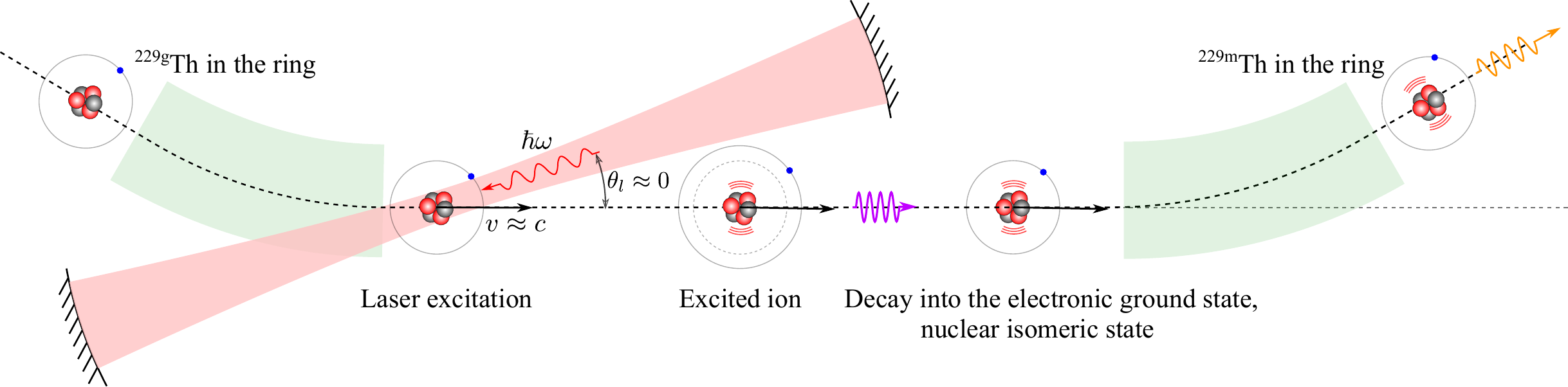}
    \caption{
    Laser excitation of $^{229m}$Th isomers in the storage ring. The laser pulse nearly counter propagates with the relativistic highly charged thorium ions. The intermediate state of the excited ion depends on the specific excitation scheme.  The intermediate state shown here corresponds to exciting the 279\,eV transition in Li-like $^{229}$Th ions, see Sec.\,\ref{subsubsec:exc_279eV}. The purple arrow denotes the photon emitted from the $^2P_{1/2}\rightarrow {}^2S_{1/2}$ electronic decay, after which the ion remains in the nuclear isomeric state. Produced $^{229m}$Th will circulate in the storage ring, with the radiative decay indicated by the orange arrow.
    }
    \label{fig:229Th_exc}
\end{figure*}

Excitation of $^{229m}$Th could be pursued via using relativistic $^{229}$Th ions in high-energy storage rings and lasers with wavelengths within the visible range or longer;
see Fig.\,\ref{fig:229Th_exc}.
In the following, we assume dealing with ultrarelativistic ($\beta\approx1$)
and highly charged ions (HCI), from Li-like $^{229}$Th ions to bare $^{229}$Th nuclei.
Activities of ground-state $^{229}$Th nuclei would be negligible due to the long half-life, relativistic time dilation together with the rather small number of $^{229}$Th ions stored in the ring.
We note that, heavy-ion storage rings have successfully been employed for studying deexcitation of nuclear excited states in highly charged ions \cite{Litvinov-2003,Reed-2010,Sun-2010,Zeng-2017}.

\subsection{Resonant excitation of the isomer}
\label{Subsec:reso_exc}

Assuming nearly head-on collisions between laser photons and the ions, due to the relativistic Doppler effect, the photon energy in the ion frame becomes \cite{Budker2021GF_nucl}
\begin{equation}
\label{Eq:energy_relation}
    \hbar\omega'=(1+\beta\cos{\theta_l})\gamma \hbar\omega\approx (1+\beta)\gamma\hbar\omega\approx 2\gamma \hbar\omega\,,
\end{equation}
where $\theta_l\approx0$ is the incident angle and $\hbar\omega$ is the laser-photon energy ($\hbar$ is the reduced Planck constant), see Fig.\,\ref{fig:229Th_exc}.
The last relation in Eq.\,\eqref{Eq:energy_relation} holds for $\beta\approx1$, i.e., $2\gamma^2\gg1$.
We use unprimed quantities ($\theta_l$ and $\hbar\omega$ here) for the lab frame, and primed quantities for the ion frame, unless otherwise specified.

Resonant excitation requires $\hbar\omega'=E_\textrm{iso}$.
The width of the photon absorption spectrum in our case is dominated by the Doppler width, $\Gamma_D\approx\hbar\omega'\Delta\gamma/\gamma$, due to the relative ion-energy spread $\Delta\gamma/\gamma$. Since $\Delta\gamma/\gamma$ is typically smaller than the relative uncertainty of the isomeric-state energy, we need to scan across the possible energy range of the isomeric state to search for the resonance. This can be done by tuning $\gamma^{\textrm{ave}}$ (the central value of $\gamma$) of the ion bunch and/or the laser-photon energy, which is discussed in Sec.\,\ref{Sec:search_res}. We assume below $2\gamma^{\textrm{ave}}\cdot\hbar\omega=E_\textrm{iso}$.


We calculate the population transfer using the density matrix approach in the ion frame, solving the master equation \cite{Buervenich2006}
\begin{equation} \label{Eq:DM_TLS} 
    \frac{d\rho}{dt} =
    \frac{1}{i\hbar}[H_{\textrm{eff}},\rho] 
    - \frac{\gamma_{\textrm{SE}}}{2}
    \begin{pmatrix}
    -2\rho_{ee} & \rho_{ge}\\
    \rho_{eg} & 2\rho_{ee}
    \end{pmatrix},
\end{equation}
where the interaction-picture Hamiltonian $H_{\textrm{eff}}$ in the rotating wave approximation  is given by 
\begin{equation}
    H_{\textrm{eff}} = -\hbar\begin{pmatrix}
    0 & \frac{\Omega_R(t)}{2}\\
    \frac{\Omega_R(t)}{2} & \Delta
    \end{pmatrix},
\end{equation}
with $\Omega_R(t)$ being the time-dependent Rabi frequency, and $\Delta=2\gamma\omega-E_{\textrm{iso}}/\hbar=(\gamma/\gamma^{\textrm{ave}}-1)E_{\textrm{iso}}/\hbar$ being the detuning for taking account of the ion energy spread.
$\gamma_{\textrm{SE}}=\Gamma_{\textrm{rad}}/\hbar$ is the spontaneous decay rate of the excited state.
We assumed employing fully coherent laser pulses.

To resonantly excite a large fraction of the ions,
we can tune the laser spectral linewidth in the ion frame to
be close to the Doppler width due to ion-energy spread; see Ref.\,\cite{Bieron2022OpticalPump}. This can be achieved by adjusting the duration of the laser pulse. The pulse is assumed to have a Gaussian temporal profile in intensity
$I'(t)=I_{0}'\exp{[-(4\ln{2}) t^2/t_\textrm{p}'^2]}$, where
$I_{0}'$ is the peak intensity
and $t_\textrm{p}'$ is the full width at half maximum (FWHM in intensity) pulse duration in the ion frame. The corresponding spectral width of the laser light
in the ion frame is $\Gamma_\textrm{p}'\approx (4\ln{2})\hbar/t_\textrm{p}'$.
$\Gamma_\textrm{p}'\approx\Gamma_D$ leads to 
\begin{equation} \label{Eq:pulse_dur}
    t_\textrm{p}'\approx4\ln{(2)}\frac{\gamma}{\Delta\gamma}\frac{1}{\omega'}\,.
\end{equation}

Alternatively, we can introduce incident-angle divergence of the laser beam to match the Doppler width as frequently done in experiments with atomic beams; see, for example, Ref.\,\cite{Nguyen2000DivLas} and references therein.
However, this might make the approximation of nearly head-on collisions between the laser beam and ion bunch invalid.
Therefore, in the following we consider head-on collisions and tuning the pulse duration of Gaussian shaped laser pulses according to Eq.\,\eqref{Eq:pulse_dur} to match the laser-pulse spectrum with the photon absorption spectrum considering the ion energy spread.

During such pulse duration the radiative decay of the excited state is negligible, i.e., $\gamma_{\textrm{SE}}\cdot t_\textrm{p}'\ll1$. Therefore, Eq.\,\eqref{Eq:DM_TLS}
can be simplified into  
\begin{equation} \label{Eq:DM_TLS_v2} 
    \frac{d\rho}{dt} =
    \frac{1}{i\hbar}[H_{\textrm{eff}},\rho]\,.
\end{equation}

The time-dependent Rabi frequency is
\begin{equation}
\label{Eq:Rabi_red}
    \Omega_{\textrm{R}}(t)=\frac{\mu B'(t)}{\hbar}\approx\left(\frac{6\pi c^2 \Gamma_\textrm{rad} I'}{\hbar^2\omega'^3}\right)^{1/2}\,,
\end{equation}
where $\mu$ is the magnetic dipole ($M1$) transition moment, related to the radiative width of the isomeric state, and $B'=(2\mu_0 I'/c)^{1/2}$ is the magnetic-field amplitude of the laser light with $\mu_0$ being the vacuum permeability.

The population-transfer fraction induced by one laser pulse is dependent on the detuning, denoted as $\rho_{ee}(\Delta)$. The average population-transfer fraction can be estimated as
\begin{equation} \label{Eq:ave_popu_tran}
    \rho_0 = \int \frac{1}{\sqrt{2\pi}\sigma}\exp{\left(-\frac{\Delta^2}{2\sigma^2}\right)} \rho_{ee}(\Delta) d\Delta
\end{equation}
by averaging over the whole ion ensemble.
Here, we assume that the detuning due to the ion energy spread has a Gaussian distribution with the FWHM being $\Gamma_D/\hbar$, so that $2(2\ln{2})^{1/2}\sigma =\Gamma_D/\hbar$. We consider adjusting the laser beam waist $w_l$ to match the 1-$\sigma$ radius $w_b$ of the ion bunch, leading to $S_l\approx S_b$, where $S_{l(b)}=\pi w_{l(b)}^2$ are the transverse cross sections of the laser beam and ion bunch, respectively.
The number of ions excited by a single laser pulse is about $\rho_0\cdot N_{\textrm{ion}}$ with $N_\textrm{ion}$ being the number of ions per bunch.

Implementing a $\pi$ pulse [$\int\Omega_{\textrm{R}}(t)dt=\pi$] for ions in the resonant group ($\Delta\approx0$) requires a peak laser-pulse intensity of 
\begin{equation} \label{Eq:Intensity_pi}
    I_{0\pi}'= \frac{\ln{2}}{3}\frac{\hbar^2\omega'^3}{t_\textrm{p}'^2c^2\Gamma_\textrm{rad}}\approx \frac{1}{48\ln{(2)}}\frac{\hbar^2\omega'^5}{c^2\Gamma_{\textrm{rad}}}\left(\frac{\Delta\gamma}{\gamma}\right)^2\,,
\end{equation}
corresponding to a laser-pulse energy of $E_{\textrm{p}}'\approx I_0'\cdot t_\textrm{p}'\cdot S_b$.
These laser-pulse parameters are for the ion frame and are related to the lab-frame parameters by
\begin{equation} \label{Eq:para_frame}
\begin{split}
  t_\textrm{p}'&=t_\textrm{p}/2\gamma\, ,\\
  E_{\textrm{p}}'&=2\gamma E_{\textrm{p}}\, , \\
  I'& =(2\gamma)^2 I\,.
\end{split}
\end{equation}

The $\pi$-pulse intensity in Eq.\,\eqref{Eq:Intensity_pi} would be high when having
a narrow radiative width and a large relative ion-energy spread. Therefore, we instead consider isomer excitation in $^{229}$Th nuclei using repeated pumping, seen by an ion over multiple round trips (Sec.\,\ref{subsubsec:reso_multi_pulse}). Such excitation relaxes the requirements on individual laser pulses via using a longer excitation sequence.
We will discuss the possibility of excitation using $\pi$ pulses in H- or Li-like $^{229}$Th ions (Sec.\,\ref{subsubsec:res_H-like}), since the expected NHM effect increases the radiative width by orders of magnitude compared to bare nuclei.

\subsubsection{Excitation in bare nuclei using repeated pumping}
\label{subsubsec:reso_multi_pulse}

To realize repeated pumping, we need the same repetition frequency and proper synchronization of laser pulses and ion bunches.
Thus, the laser-pulse repetition rate is adjusted to be $f_l=N_b\cdot f_b$, with $N_b$ and $f_b$ being the number of ion bunches and the circulating frequency of a single ion bunch in the storage ring, respectively.

The population-transfer fraction in each ion bunch, denoted as $\rho_{\textrm{ex}}$, depends on the number of laser pulses having collided with the ion bunch, denoted as $n_{\textrm{col}}$.
The population evolution under repeated pumping ($\rho_{\textrm{ex}}$ versus $n_{\textrm{col}}$) can be obtained by repeatedly calculating Eqs.\,\eqref{Eq:DM_TLS_v2} and \eqref{Eq:ave_popu_tran} in a loop [Eq.\,\eqref{Eq:ave_popu_tran} resets initial
conditions of population for Eq.\,\eqref{Eq:DM_TLS_v2}].

\begin{figure}[!htpb]\centering
    \includegraphics[width=0.95\linewidth]{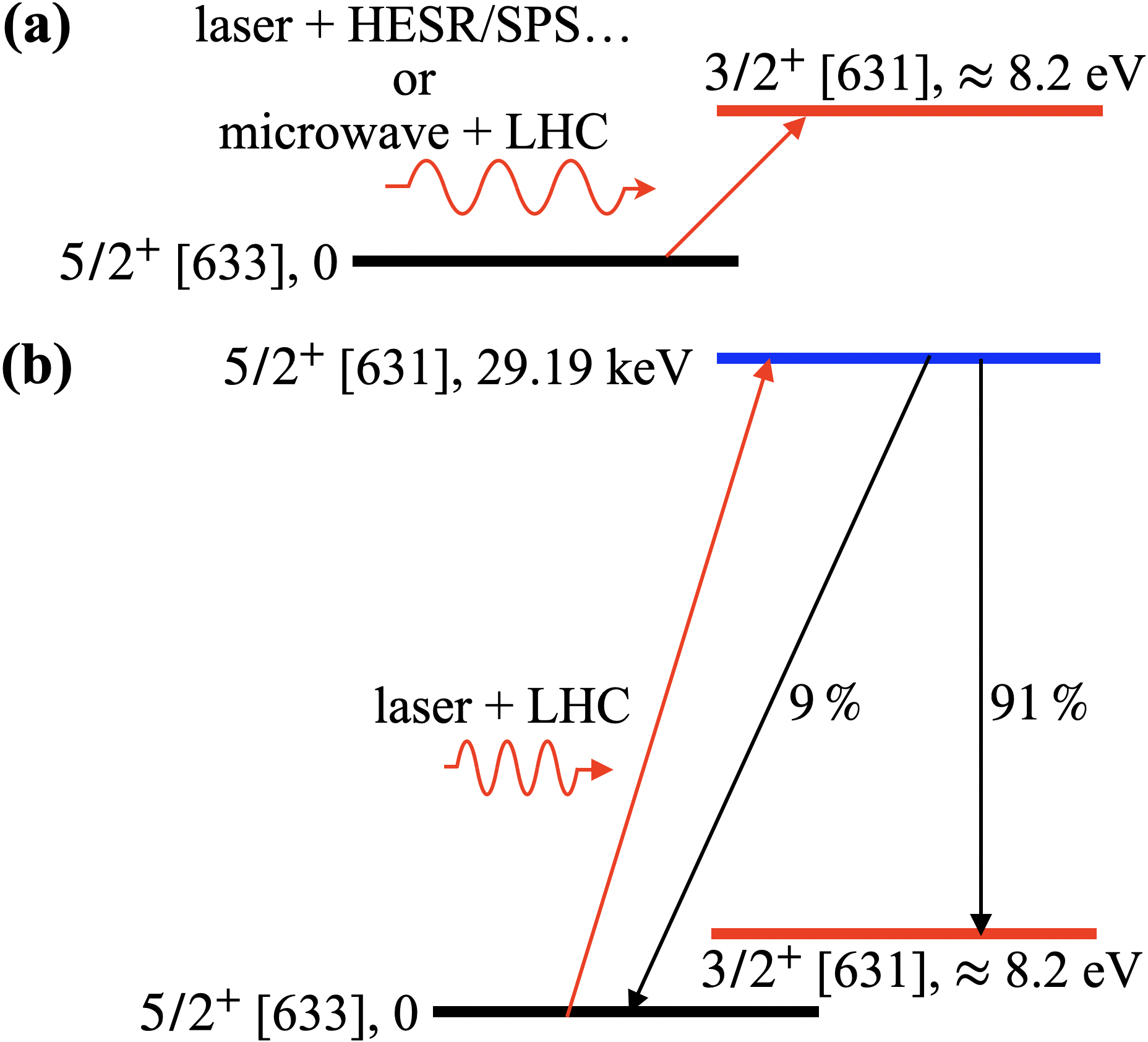}
    \caption{Production of $^{229m}$Th isomers through (a) resonant excitation or (b) excitation via the second nuclear excited state.
    The second scenario requires a large Lorentz factor of the ion bunch, which could be achieved at the Large Hadron Collider (LHC).
    }
    \label{fig:Th_excitation}
\end{figure}

To  give  a  specific  example, we consider exciting relativistic $^{229}$Th ions at the SPS; see also Fig.\,\ref{fig:Th_excitation}\,(a) and Table\,\ref{tab:ion-beam_facilities}.
There can be up to $N_b=36$ ion bunches circulating in the storage ring with $S_b\approx\pi(0.5\textrm{mm})^2$ and $N_{\textrm{ion}}\approx10^8$ per bunch.
A pulsed CO$_2$ laser ($\hbar\omega=0.117\,$eV) having an energy of $E_{\textrm{p}}=100\,\mu$J per pulse and operating at a repetition rate of $f_l=1.56\,$MHz synchronized to the circulating ion bunches can be used to excite the nuclear isomeric state in $^{229}$Th ions.
The required Lorentz factor of the ion bunch is $\gamma\approx35$ ($\Delta\gamma/\gamma\approx10^{-4}$ at the SPS).
We further assume using a Fabry–P\'erot cavity offering a factor of $\approx10^5$ enhancement of the laser power inside the cavity, leading to $E_{\textrm{p}}=10\,$J.
The average power is
$E_\textrm{p}\cdot f_l = 1.56\times10^7\,$W.

\begin{figure}[!htpb]\centering
    \includegraphics[width=0.9\linewidth]{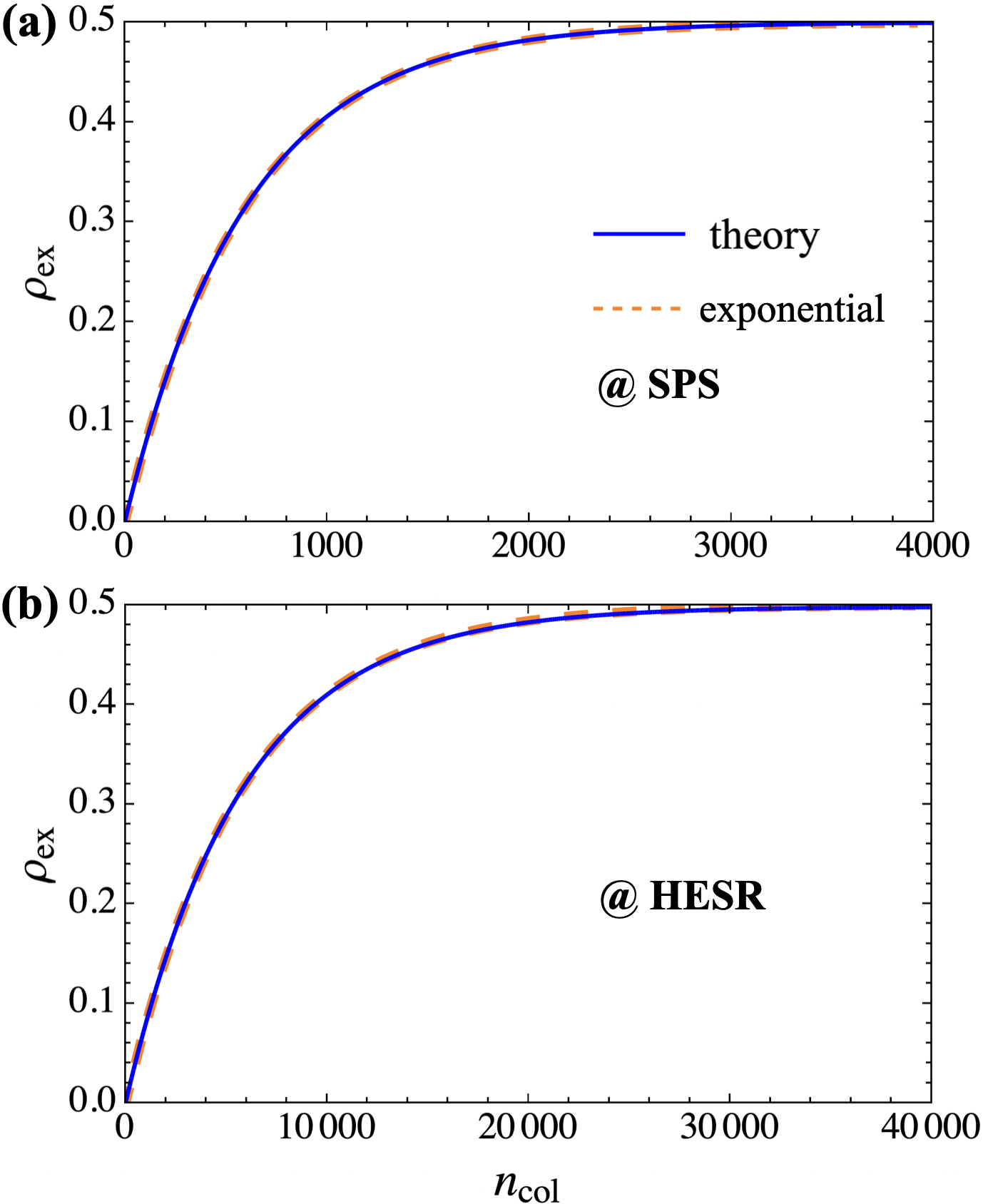}
    \caption{
    Repeated pumping at (a) SPS and (b) HESR. The blue solid curve is derived from repeatedly calculating the population transfer using Eqs.\,\eqref{Eq:DM_TLS_v2} and  \eqref{Eq:ave_popu_tran}, which also agrees with Eq.\,\eqref{Eq:exp_behavior} shown as the orange dashed curve.
    The characteristic time for reaching population saturation is about $60\,$ms in both cases, see the text.
    }
    \label{fig:Rep_pump_SPS_HESR}
\end{figure}

We obtain $t_\textrm{p}'\approx2.2\,$ps and $I_0'\approx4.0\times10^{20}\,$W/m$^2$. The $I_0'$ is orders of magnitude below the intensity required for a $\pi$-pulse $I_{0\pi}'\approx7.9\times10^{23}\,$W/m$^2$ [see Eq.\,\eqref{Eq:Intensity_pi}]. 
The numerical results show
$\rho_{ee}(\Delta=0)\approx1.25\times10^{-3}$ and  $\rho_0\approx8.9\times10^{-4}$; see Table\,\ref{tab:repeat_pump} (for lab-frame parameters).
This indicates that the ion-energy spread leads to a relatively small correction [$\rho_0\approx0.71\cdot\rho_{ee}(\Delta=0)$], which is as expected from matching the laser spectrum with the photon-absorption spectrum of ions.
The population transfer is presented in Fig.\,\ref{fig:Rep_pump_SPS_HESR}\,(a).
The maximal population-transfer fraction is $\rho_{\textrm{max}}\approx0.50$. $\rho_{\textrm{ex}}=0.99\rho_{\textrm{max}}$ is reached after $n_{\textrm{col}}\approx2600$ collisions, corresponding to an excitation time of $t_{\textrm{ex}}=n_{\textrm{col}}/f_b\approx60\,$ms.
We note that with $\rho_0\ll\rho_{\textrm{max}}$, the increase of $\rho_{\textrm{ex}}$ can be well described by a simple exponential behavior (see Fig.\,\ref{fig:Rep_pump_SPS_HESR})
\begin{equation} \label{Eq:exp_behavior}
    \rho_{\textrm{ex}} =  \rho_{\textrm{max}}\left[1-\exp{\left(-\frac{\rho_0}{\rho_{\textrm{max}}}n_{\textrm{col}}\right)}\right]\,.
\end{equation}
This can be understood from the fact that the excitation process, using repeated pumping, is effectively incoherent since the optical phase for consecutive interactions of an ion with the laser light is random.

We consider another similar example of resonant excitation of $^{229m}$Th isomers, this time at HESR.
Electron cooling of ion bunches with energies of up to about 3.5\,A\,GeV and variable stochastic cooling up to the highest energies will be available in the HESR \cite{Stohlker-2015b}.
As a consequence, a smaller ion-beam energy spread (down to $\Delta\gamma/\gamma\lesssim10^{-5}$) can be achieved.
Assuming using cooled relativistic ions with $\gamma\approx6$ ($\Delta\gamma/\gamma\approx10^{-5}$), the energy of laser photons should be $\hbar\omega\approx8.19/12=0.683\,$eV.
Consider that there is one ion bunch containing $\approx10^8$ $^{229}$Th ions interacting with laser pulses repeated at $f_l=522$\,kHz (synchronized to the ion bunch) and again, $E_{\textrm{p}}=10\,$J (inside a Fabry–P\'erot cavity).

Following a procedure similar to the one discussed above for SPS, we obtain the results for HESR, as shown in Table\,\ref{tab:repeat_pump}.
We have, again, $\rho_0\approx0.71\cdot\rho_{ee}(\Delta=0)$, and the increase of $\rho_{\textrm{ex}}$ can be well described by Eq.\,\eqref{Eq:exp_behavior}; see Fig.\,\ref{fig:Rep_pump_SPS_HESR}\,(b).

Direct resonant excitation could also be pursued at the LHC, where the large $\gamma$ could allow excitation using the magnetic field produced in an undulator (e.g., a Halbach array made of permanent magnets) or a microwave source \cite{Budker2021GF_nucl}.

\begin{table}[htpb]
    \centering
    \begin{tabular*}{\linewidth}{@{\extracolsep{\fill}} lcccc cc}
    \hline 
    \hline
    Facility    &$\gamma$    &$t_\textrm{p}\,$(ps)    &$I_0\,$(W/m$^2$)  &$\rho_{ee}(\Delta=0)$   &$\rho_0$   &$t_\textrm{ex}\,$(ms) \\
    \hline \\[-0.2cm]
    SPS       &35   &150   &$8.2\times10^{16}$  &$1.25\times10^{-3}$   &$8.9\times10^{-4}$   &60\\
    HESR      &6    &260  &$2.7\times10^{15}$   &$1.22\times10^{-4}$  &$8.6\times10^{-5}$   &56\\
    \hline
    \hline
    \end{tabular*}
    \caption{
    Repeated pumping of $^{229m}$Th isomers at the SPS and HESR. $\gamma$, $t_\textrm{p}$ and $I_0$ are some of the considered parameters of the ion bunch and laser beam. $\rho_{ee}(\Delta=0)$ and $\rho_0$ are the population-transfer fraction of the ion group resonant with the laser pulse and the transfer fraction averaged over all ions, respectively, in the first round.
    $t_\textrm{ex}$ is the time needed to achieve $99\%$ of the maximal population transfer, $\rho_{\textrm{max}}\approx0.5$ in both cases (corresponding to $5\times10^7$ ions per bunch).
    }
    \label{tab:repeat_pump}
\end{table}

\subsubsection{Excitation in H- or Li- like Th ions using $\pi$ pulses}
\label{subsubsec:res_H-like}

In H- or Li-like Th ions, the lifetime of the isomeric state could be reduced by orders of magnitude compared to bare nuclei because of the nuclear hyperfine mixing (NHM) effect; see, for example, Refs.\,\cite{Karpeshin1998NHM,Tkalya2016_NHM,Shabaev2021NHM}.
Direct resonant excitation of $^{229m}$Th isomers using H- or Li-like $^{229}$Th ions was proposed at GSI; see, for example, Ref.\,\cite{Brandau2020_NHM_Th}.
The increased radiative width would significantly reduce the peak laser-pulse intensity required for implementing a $\pi-$pulse.
We calculate below the transition dipole moment considering the NHM effect and the laser-pulse intensity needed for isomer excitation using a $\pi-$pulse.
We show that, with the NHM effect, Eq.\,\eqref{Eq:Intensity_pi} still works as a proper estimate, and the required intensity could be experimentally achievable.

\begin{figure}[!htpb]\centering
    \includegraphics[width=\linewidth]{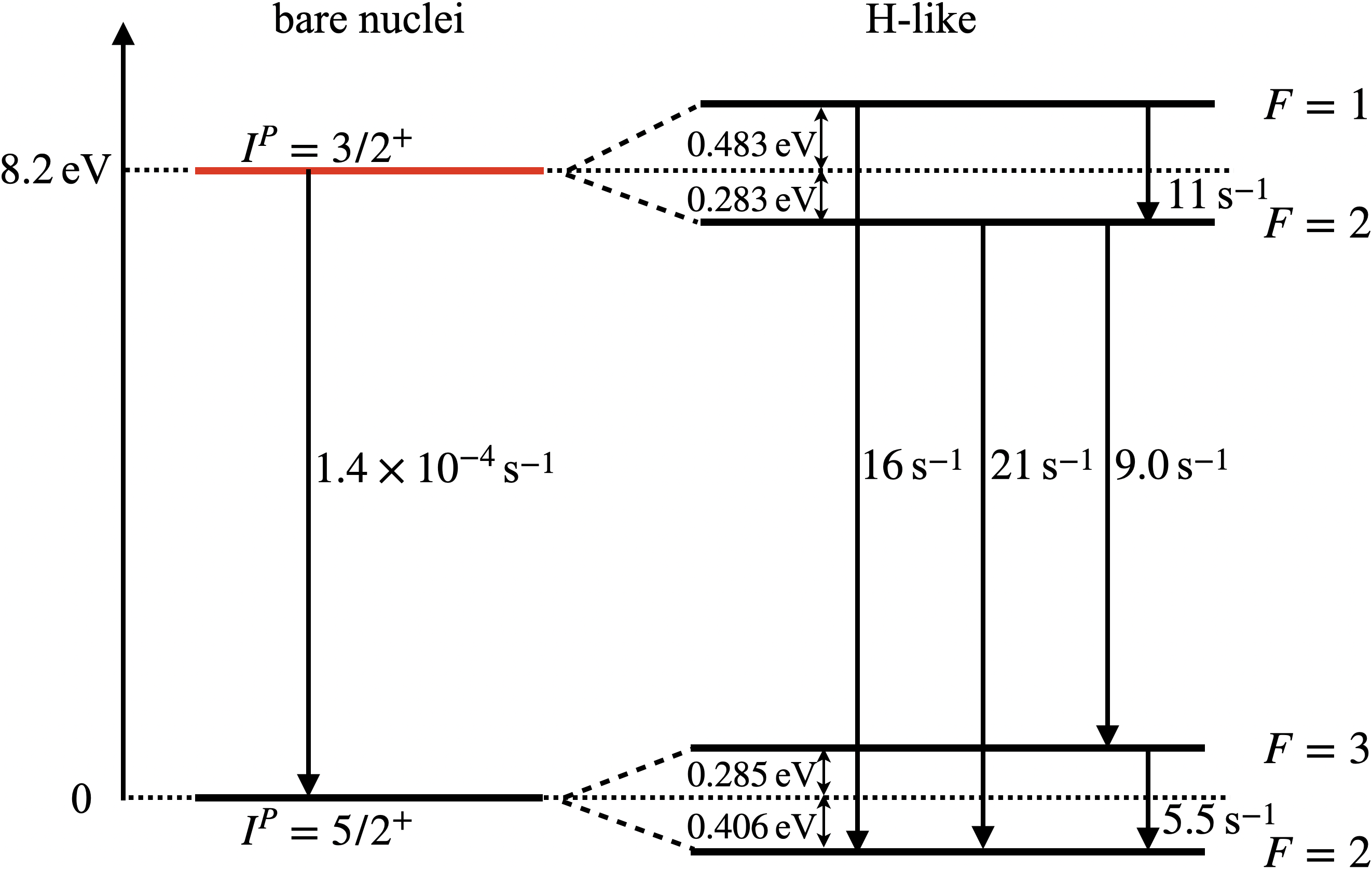}
    \caption{
    Energy levels of the nuclear ground and isomeric states and related radiative transition rates ($\Gamma_{\textrm{rad}}/\hbar$) in bare thorium nuclei (left) and H-like thorium ions (right). The values are from Ref.\,\cite{Shabaev2021NHM}.
    Due to the NHM effect in H-like thorium, the two $F=2$ HFS levels are mixed, leading to significantly higher radiative transition rates from the isomeric state to the ground state compared to bare nuclei. Similar effect also exists in Li-like thorium ions; see the text.
}
    \label{fig:H-like_Th}
\end{figure}

The electronic ground state of H-like $^{229}$Th ions, $^2S_{1/2}$, has two hyperfine components, $F=2,3$ ($F=1,2$) for nuclei in the ground (isomeric) state.
$F$ is the total angular momentum quantum number.
NHM mixes states with the same $F$ and the same magnetic quantum number $M_F$.
The two $F=2$ levels of different nuclear states ($3/2^+$ and $5/2^+$) can be represented as \cite{Shabaev2021NHM}
\begin{align} \label{Eq:NHM}
    &\overline{\ket{\frac{5}{2}^+,F, M_{F}}}=\sqrt{1-b^2}\ket{\frac{5}{2}^+,F, M_{F}}-b\ket{\frac{3}{2}^+,F, M_{F}}\nonumber\\
    &\overline{\ket{\frac{3}{2}^+,F, M_{F}}}=\sqrt{1-b^2}\ket{\frac{3}{2}^+,F, M_{F}}+b\ket{\frac{5}{2}^+,F, M_{F}}\,,
\end{align}
where $b$ is the NHM coefficient.
We use below $b=-2.9\times10^{-2}$ and $-4.9\times10^{-3}$ for H- and Li-like thorium ions, respectively, according to the theoretical calculations in Ref.\,\cite{Shabaev2021NHM}.
Theoretical results of the hyperfine structure (HFS) energies and radiative transition rates between the HFS levels are presented in Ref.\,\cite{Shabaev2021NHM} and also shown in Fig.\,\ref{fig:H-like_Th}.

The higher HFS level of H-like $^{229g}$Th (``g" denotes the nuclear ground state)
has a lifetime of about 183\,ms in the ion frame.
Therefore, after $^{229g}$Th ions are produced and circulate in the storage ring for a sufficiently long time, all the ions are in the lower HFS level, $\overline{\ket{5/2^+,F=2}}$.
Interaction of the ion beam with the magnetic fields of the ring lattice and collisions with rest gas atoms and electrons of the cooler may alter the population of the hyperfine states. However, no re-population of the upper HFS was so far observed in sensitive HCI experiments at the ESR\,\cite{Seelig-1998, Litvinov-2007,Winckler-2009, Atanasov-2012, Kienle-2013, Ozturk-2019}. Therefore, in the following we neglect the machine-induced (de)excitation of hyperfine states.

We consider exciting the transition from  $\ket{g}=\overline{\ket{5/2^+,F=2}}$ to $\ket{e}=\overline{\ket{3/2^+,F=2}}$.
The radiative width of the corresponding decay channel $\ket{e}\rightarrow\ket{g}$
is $\Gamma_\textrm{rad}\approx1.4\times10^{-14}\,$eV, and
the transition energy is $\hbar\omega'=8.31(13)\,$eV \cite{Shabaev2021NHM}.
Again, as discussed above, we adjust the laser pulse duration according to Eq.\,\eqref{Eq:pulse_dur} and search for the resonance, so that we have $\Gamma_\textrm{p}'\approx\Gamma_D$ and $2\gamma^{\textrm{ave}}\cdot\hbar\omega\approx8.31\,$eV.

The dipole moment of the $M1$ transition between sublevels $\overline{\ket{5/2^+,F=2,M_F}}$ and $\overline{\ket{3/2^+,F=2,M_F'}}$ is
\begin{equation} \label{Eq:dipole_NHM_res}
\begin{split}
\mu&=\left| \overline{\bra{\frac{3}{2}^+,F, M_{F}'}}M_{L\lambda} \overline{\ket{\frac{5}{2}^+,F, M_{F}}} \right|\\
&\approx \left| \overline{\bra{\frac{3}{2}^+,F, M_{F}'}}M_{L\lambda}^{(e)} \overline{\ket{\frac{5}{2}^+,F, M_{F}}} \right|
\\
&= \left|\braket{FM_FL\lambda|FM_F'}\right| \frac{|b|\sqrt{1-b^2}}{\sqrt{2F+1}}
\\
&\Big|
\bra{\frac{5}{2}^+,F}|M^{(e)}_{L}|\ket{\frac{5}{2}^+,F}-\bra{\frac{3}{2}^+,F}|M^{(e)}_{L}|\ket{\frac{3}{2}^+,F}\Big|\,,
\end{split}
\end{equation}
where
$\braket{FM_FL\lambda|FM_F'}$ is the Clebsch–Gordan coefficient,
$\braket{\cdot||M_L||\cdot}$ is the reduced matrix element,
$M_{L\lambda}=M_{L\lambda}^{(e)}+M_{L\lambda}^{(n)}$ is the magnetic multipole operator composed of the electronic part $M_{L\lambda}^{(e)}$ and the nuclear part $M_{L\lambda}^{(n)}$ with $L$ and $\lambda$ representing the radiative transition multipolarity and photon polarization. Here, we have $L=1$ for the $M1$ transition. $\mu$ is dominated by the electronic part as a result of the NHM,
which can be seen from the fact that
$|b|\gg \mu_N/\mu_B\approx5\times10^{-4}$
($\mu_N$ and $\mu_B$ are the nuclear and Bohr magneton, respectively).

$\mu$ can be derived from the radiative rate of the $\ket{e}\rightarrow\ket{g}$ transition  \cite{Shabaev2021NHM}:
\begin{equation} \label{Eq:rad_rate}
\begin{split}
    \frac{\Gamma_{\textrm{rad}}}{\hbar}&= \frac{\mu_0}{3\pi\hbar}\left(\frac{\omega'}{c}\right)^3\frac{1}{2F+1}\\
    &\sum_{M_{F}',M_{F},\lambda}
    |\overline{\bra{\frac{5}{2}^+FM_{F}}}
     M_{L\lambda}
    \overline{\ket{\frac{3}{2}^+FM_{F}'}}|^2\\
    &\approx \frac{\mu_0}{3\pi\hbar}\left(\frac{\omega'}{c}\right)^3
    \sum_{M_{F}',M_{F},\lambda}
    |\braket{FM_FL\lambda|FM_F'}|^2\cdot \mu_{\textrm{r}}^2\\
    &=\frac{\mu_0}{3\pi\hbar}\left(\frac{\omega'}{c}\right)^3\frac{\mu_{\textrm{r}}^2}{2F+1}\,.
\end{split}
\end{equation}
We have used the reduced magnetic dipole moment $\mu_{\textrm{r}}=(2F+1)^{1/2}\mu/|\braket{FM_FL\lambda|FM_F'}|$, which is independent of the magnetic quantum numbers ($M_{F}, M_{F}',\lambda$).

The reduced Rabi frequency $\Omega_{\textrm{r}}\sim[\mu_r/(2F+1)^{1/2}] B'/\hbar$ satisfies Eq.\,\eqref{Eq:Rabi_red}, so the peak laser-pulse intensity required in the ion frame for a (nominal) $\pi-$pulse, $\int\Omega_{\textrm{r}}(t)dt=\pi$, can be deduced using Eq.\,\eqref{Eq:Intensity_pi}.
Such intensity would produce $\rho_{ee}(\Delta=0)\approx1$. Considering the ion-energy spread, the average excitation fraction is $\rho_0\approx0.7\cdot\rho_{ee}(\Delta=0)\approx0.7$; see Eq.\,\eqref{Eq:ave_popu_tran}.

We return to the example of exciting $^{229m}$Th at the SPS using relativistic ion bunches with $\gamma\approx35$ ($\Delta\gamma/\gamma\approx10^{-4}$) but for now dealing with H-like $^{229}$Th ions. The required laser-beam parameters (in the lab frame) for implementing the $\pi-$pulse are
$I_{0}=1.1\times10^{15}$\,W/m$^2$ and
$E_\textrm{p}= 140\,$mJ
($t_\textrm{p}\approx150\,$ps).
For the HESR, we derive
$I_{0}=3.7\times10^{14}$\,W/m$^2$ and $E_\textrm{p}\approx 1.4\,$J ($t_{\textrm{p}}\approx260\,$ps).
The pulse energies required here are smaller than $E_\textrm{p}=10$\,J assumed in Sec.\,\ref{subsubsec:reso_multi_pulse}, thus easier to achieve experimentally. Also, when exciting isomers using $\pi-$pulses, we do not need high repetition rate of the laser pulses.
For Li-like $^{229}$Th ions,
the NHM coefficient $b$ is smaller compared to H-like $^{229}$Th ions, leading to  lower radiative decay rates of the HFS levels \cite{Shabaev2021NHM}.

The higher HFS level of Li-like $^{229g}$Th in the electronic ground state $1s^22s_{1/2}$ ($^2S_{1/2}$)
has a long lifetime of about 27\,s in the ion frame.
Therefore, instead of considering all ions decaying into the lower HFS level, we can assume that both HFS levels ($\overline{\ket{5/2^+,F=2}}$ and $\ket{5/2^+,F=3}$) are populated and the relative population is proportional to the degeneracy factor $g_{F}=2F+1$. 
In this case, besides exciting transitions from $\overline{\ket{5/2^+,F=2}}$, we also need to consider transitions from
$\ket{5/2^+,F=3}$ to $\overline{\ket{3/2^+,F=2}}$.
We note that
Eq.\,\eqref{Eq:Rabi_red}
also holds for those transitions.
The radiative transition rates ($\Gamma_{\textrm{rad}}/\hbar$) of $\overline{\ket{3/2^+,F'=2}}\rightarrow\overline{\ket{5/2^+,F=2}}$, $\overline{\ket{3/2^+,F'=2}}\rightarrow\ket{5/2^+,F=3}$
and $\ket{3/2^+,F'=1}\rightarrow\overline{\ket{5/2^+,F=2}}$
are about 0.72, 0.39, and 0.45\,s$^{-1}$ \cite{Shabaev2021NHM},
respectively.
The required laser intensity for a (nominal) $\pi-$pulse corresponding to those transitions can be obtained using Eq.\,\eqref{Eq:Intensity_pi}.

One can also produce polarized Li-like $^{229g}$Th ions via optical pumping using the electronic transition between $1s^{2}2s_{1/2}$ ($^2S_{1/2}$) and $1s^{2}2p_{1/2}$ ($^2P_{1/2}$); see Sec.\,\ref{Subsec:elec_tran},
from which fully polarized ions, population only in the $\ket{5/2^+,F=3,M_F=3}$ state, could be obtained. Afterwards, a $\pi-$pulse could be applied to transfer the population into the isomeric state $\overline{\ket{3/2^+,F=2,M_F=2}}$. The peak laser-pulse intensity required for $\int\Omega_{\textrm{R}}(t)dt=\pi$ ($\Omega_{\textrm{R}}=\Omega_{\textrm{r}}|\braket{331-1|22}|$) is $I_{0}\approx8.5\times10^{16}$W/m$^2$, corresponding to $E_{\textrm{p}}\approx11\,$J, at the SPS.

\subsection{Excitation via the electronic excited state in Li-like $^{229}$Th}
\label{Subsec:exc_271eV}

\subsubsection{
Forbidden $E1$ transitions opened by the NHM effect
}

\begin{figure}[!htpb]
    \centering
    \includegraphics[width=\linewidth]{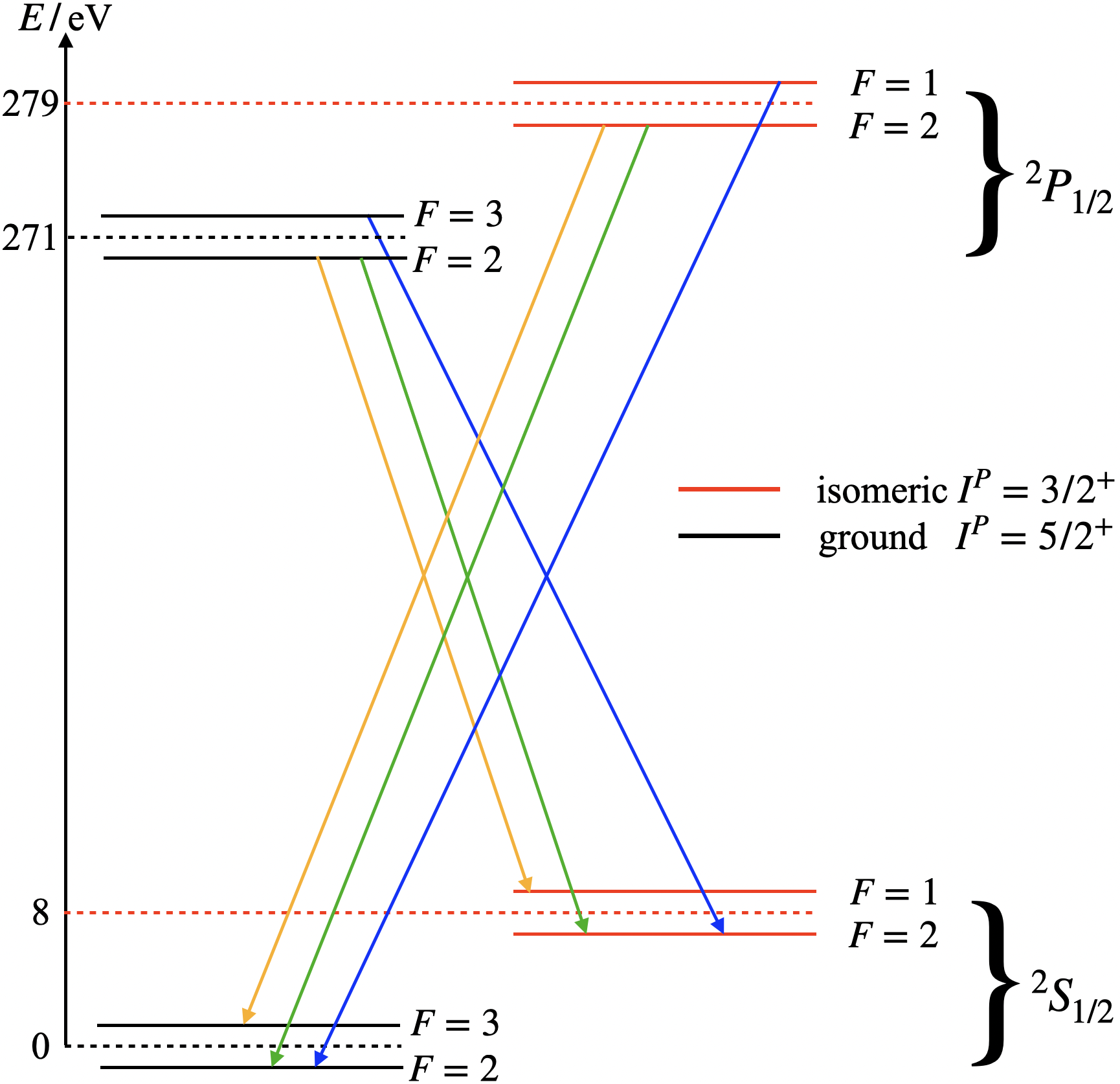}
    \caption{
    The low-lying energy levels in Li-like ${}^{229}$Th ions and transitions between hyperfine components that change both the nuclear and electronic states.
    The transitions denoted by the solid arrows, although occurring between different nuclear states, are all $E1$ transitions, due to the NHM effect. Since the mixing only occurs between states with the same total angular momentum quantum number $F$, the transitions represented by the blue (orange) arrows are allowed due to mixing of the two $F=2$ levels in the electronic ground (excited) state, ${}^2S_{1/2}$ (${}^2P_{1/2}$).  
    The transitions denoted by green arrows include contributions from mixing in both ${}^2S_{1/2}$ and ${}^2P_{1/2}$ manifolds.
    }
    \label{Fig:Li-like_Th}
\end{figure}

We present in Fig.\,\ref{Fig:Li-like_Th} the low-lying energy levels in Li-like ${}^{229}$Th ions and transitions between hyperfine components that change both the electronic and nuclear states.
Naively, one would consider those transitions to be $E1$ forbidden. However, they are actually all
$E1$ allowed due to the NHM effect, i.e., hyperfine levels in the ${}^2S_{1/2}$ (or ${}^2P_{1/2}$) with different nuclear states can be mixed.

First, consider the transitions from the nuclear ground state to the isomeric state, denoted by the arrows pointing from the left top to right bottom in Fig.\,\ref{Fig:Li-like_Th}. We denote the higher and lower levels as
\begin{equation}
\begin{split}
    \ket{e} &=\overline{\ket{{}^2P_{1/2},I_g,F_e}}\\
            &\approx \ket{{}^2P_{1/2},I_g,F_e} - \delta_{F_e,2}\cdot b_e \ket{{}^2P_{1/2},I_e,F_e}\,,\\
    \ket{g} &=\overline{\ket{{}^2S_{1/2},I_e,F_g}}\\
            &\approx \ket{{}^2S_{1/2},I_e,F_g} + \delta_{F_g,2}\cdot b_g \ket{{}^2S_{1/2},I_g,F_g}\,,
\end{split}
\end{equation}
where $\delta_{F_e,2}\cdot b_e$ and $\delta_{F_g,2}\cdot b_g$ (both $\ll1$) are the NHM factors of the HFS levels $F_e$ in the electronic excited ${}^2P_{1/2}$ state  and $F_g$ in the electronic ground state ${}^2S_{1/2}$, respectively.
Here, $\delta_{F_{g(e)},2}$ is the Kronecker delta symbol meaning that the NHM occurs between the two $F_{g(e)}=2$ levels.
$b_g\approx-4.9\times10^{-3}$ is derived in Ref.\,\cite{Shabaev2021NHM}. However, to the best of our knowledge, $b_e$ has not been calculated yet.

In Ref.\,\cite{Shabaev2021NHM}, the authors also calculated the NHM effect for boron-like ${}^{229}$Th ions in the electronic ground state, $1s^22s^22p_{1/2}$ (${}^2P_{1/2}$). The NHM factor in this case is $b_g(\textrm{B}) \approx -1.7\times10^{-3}$.
$b_e$ of Li-like ${}^{229}$Th ions should satisfy $|b_g(\textrm{B})|< |b_e(\textrm{Li})|<|b_g(\textrm{Li})|$, thus we roughly estimate it as $b_e \approx -3\times10^{-3}$.

The reduced $E1$ transition dipole between the states is
\begin{equation} \label{Eq:red_dip_NHM}
\begin{split}
    d_{e\rightarrow g}
    &= \bra{g}|E_L^{(e)}|\ket{e} \\
    &\approx \delta_{F_g,2}\cdot b_g \bra{{}^2S_{1/2},I_g,F_g}|E_L^{(e)}|\ket{{}^2P_{1/2},I_g,F_e}\\
    &\quad  -\delta_{F_e,2}\cdot b_e \bra{{}^2S_{1/2},I_e,F_g}|E_L^{(e)}|\ket{{}^2P_{1/2},I_e,F_e}\,,
\end{split}
\end{equation}
where $\bra{{}^2S_{1/2},I,F_g}|E_L^{(e)}|\ket{{}^2P_{1/2},I,F_e}$ can be further related to the reduced matrix element between the fine-structure levels according to \cite{Auzinsh2010LightAtom}
\begin{equation}
\begin{split}
    \bra{\xi_g,J_g,I,F_g}|&E_L^{(e)}|\ket{\xi_e,J_e,I,F_e} = (-1)^{J_g+I+F_e+1}\cdot\\
    &(g_{F_e}g_{F_g})^{1/2}\begin{Bmatrix}
    J_g & F_g &I\\
    F_e & J_e &1 
    \end{Bmatrix}\bra{\xi_gJ_g}|E_L^{(e)}|\ket{\xi_eJ_e}\,.
\end{split}
\end{equation}
Here, $J_{g(e)}=1/2$ is the total electronic angular momentum of the ${}^2S_{1/2}({}^2P_{1/2})$ fine structure level, $\xi_{g(e)}$ incorporates all other electronic quantum numbers, and $g_{F_g(F_e)}$ is the degeneracy factor of the $F_g(F_e)$ HFS level.
Therefore, Eq.\,\eqref{Eq:red_dip_NHM} can be rewritten as
\begin{equation} \label{Eq:red_dip_NHM_New}
\begin{split}
    \frac{d_{e\rightarrow g}}{d_{\xi_eJ_e\rightarrow\xi_gJ_g}}
    &\approx \delta_{F_g,2}\cdot (5g_{F_e})^{1/2}b_g \begin{Bmatrix}
    1/2 & 2 &5/2\\
    F_e & 1/2 &1 
    \end{Bmatrix}\\
    &\,\,+\delta_{F_e,2}\cdot (5g_{F_g})^{1/2}b_e \begin{Bmatrix}
    1/2 & F_g &3/2\\
    2 & 1/2 &1 
    \end{Bmatrix}\,,
\end{split}
\end{equation}
using the notation $d_{\xi_eJ_e\rightarrow\xi_gJ_g} = \bra{\xi_gJ_g}|E_L^{(e)}|\ket{\xi_eJ_e}$ and neglecting the overall phase factor $(-1)^{F_e}$.

Since $E1$ transitions have a radiative width \cite{Budker2004AP,Auzinsh2010LightAtom}
\begin{equation} \label{Eq:E1_dipole}
\Gamma_{a\rightarrow b}
    = \frac{\omega_{ab}^3}{3\pi\epsilon_0 c^3}\cdot \frac{1}{g_a}\cdot|d_{a\rightarrow b}|^2\,,
\end{equation}
where $g_a$ is the degeneracy factor of state $a$,
the radiative width of the transition $\ket{e}\rightarrow\ket{g}$, denoted as $\Gamma_{eg}$, can be derived from
\begin{equation} \label{Eq:rad_width}
    \frac{\Gamma_{eg}}{\Gamma_{\xi_eJ_e\rightarrow\xi_gJ_g}} = \frac{2J_e+1}{2F_e+1}\left(\frac{\omega_{eg}}{\omega_{\xi_eJ_e\rightarrow\xi_gJ_g}}\right)^3\left| \frac{d_{e\rightarrow g}}{d_{\xi_eJ_e\rightarrow\xi_gJ_g}}\right|^2\,.
\end{equation}
 $\Gamma_{\xi_eJ_e\rightarrow\xi_gJ_g}=\hbar/T_{\xi_eJ_e}=1.12\times10^{-5}\,$eV is the radiative width of the $^2P_{1/2}\rightarrow{}^2S_{1/2}$ transition, with $T_{\xi_eJ_e}=5.90\times10^{-11}\,$s being the radiative lifetime of the $^2P_{1/2}$ state \cite{Theodosiou1991Li-Ion}.

Transitions from the nuclear excited state to ground state, denoted by the arrows pointing from the right top to left bottom in Fig.\,\ref{Fig:Li-like_Th}, can be treated similarly.
The estimated radiative widths are presented in Table\,\ref{tab:rad_width_NHM}.
These widths are larger than that of the purely nuclear radiative transition (without considering the NHM effect)  by about nine orders of magnitude.

\begin{table}[htpb]
    \centering
    \begin{tabular*}{\linewidth}{@{\extracolsep{\fill}} l cc }
    \hline 
    \hline
    Transition    & $E_{if}$\,(eV)  & $\Gamma_{if}\,$(eV)   \\
    ${}^2P_{1/2},I_i^P, F_i\rightarrow {}^2S_{1/2},I_f^P, F_f$ & &\\
    \hline \\[-0.2cm]
    $5/2^+,3\rightarrow 3/2^+,2$     &263   &$1.4\times10^{-10}$  \\
    $5/2^+,2\rightarrow 3/2^+,1$     &263   &$4.6\times10^{-11}$    \\
    $5/2^+,2\rightarrow 3/2^+,2$     &263   & $2.0\times10^{-10}$  \\
    $3/2^+,1\rightarrow 5/2^+,2$     &279   &$2.4\times10^{-10}$    \\
    $3/2^+,2\rightarrow 5/2^+,3$     &279   &$8.5\times10^{-11}$    \\
    $3/2^+,2\rightarrow 5/2^+,2$     &279   &$2.9\times10^{-10}$   \\
    \hline
    \hline
    \end{tabular*}
    \caption{
    Energies $E_{if}$ and radiative widths $\Gamma_{if}$ estimated for transitions between HFS levels in Li-like ${}^{229}$Th ions. These transitions change both the electronic and nuclear states, see also Fig.\,\ref{Fig:Li-like_Th}.
    For comparison, we note that the radiative widths of the pure electronic transition, $^2P_{1/2}\rightarrow{}^2S_{1/2}$, and the pure nuclear transition, $I_e^P=3/2^+\rightarrow I_g^P=5/2^+$ (without NHM), are about
    $1.12\times10^{-5}\,$ and $8.8\times10^{-20}\,$eV, respectively.
    }
    \label{tab:rad_width_NHM}
\end{table}

\subsubsection{
Isomer excitation via STIRAP
}
\label{subsubsec:STIRAP_Li-like}

In this Section we discuss the possibility of exciting $^{229m}$Th isomers via the intermediate electronic excited state $1s^22p_{1/2}({}^2P_{1/2})$ in Li-like thorium ions, relying on the NHM effect. This is essentially addressing a three-level system in a so-called $\Lambda$ configuration with two laser fields that transfer the population between the nuclear ground state and isomeric state. In the following we discuss coherent population transfer via STIRAP.

\begin{figure}[!htpb]
    \centering
    \includegraphics[width=\linewidth]{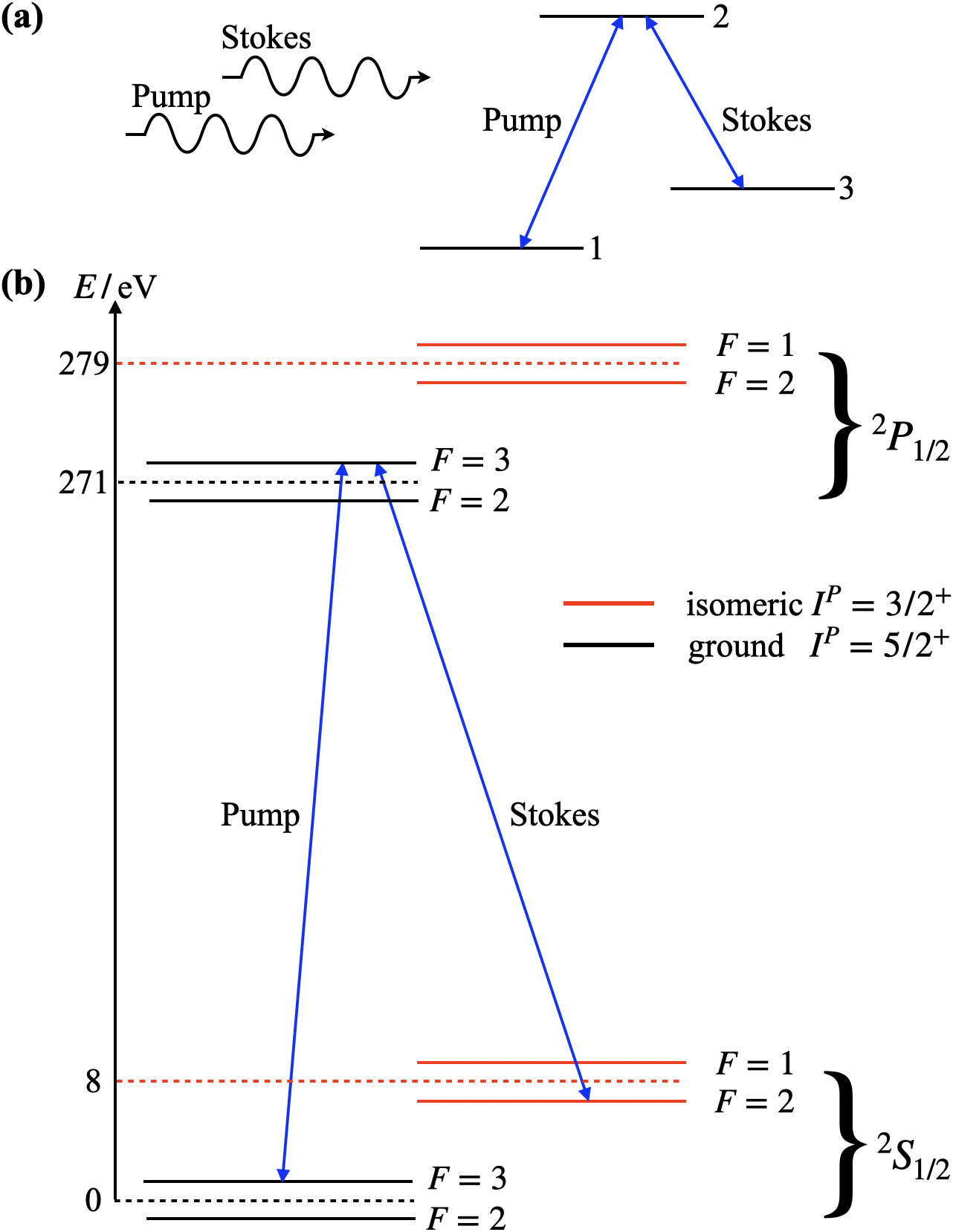}
    \caption{
    (a) A generic STIRAP scheme. The pump and Stokes and pulses interact with a three-level $\Lambda$ system. Under STIRAP conditions, the population from state 1 can be fully transferred to state 3, while the population of state 2 remains nominally zero throughout the process.
    (b) Excitation of $^{229m}$Th via the electronic state $1s^22p_{1/2}$(${}^2P_{1/2}$) in Li-like thorium. Red solid lines denote energy levels of thorium ions in the nuclear isomeric state $I^P=3/2^+$. Other energy levels are for thorium ions in the nuclear ground state $I^P=5/2^+$. Here, as an example, we consider a pump laser driving the transition between $\ket{5/2^+,{}^2S_{1/2},F=3}$ and $\ket{5/2^+,{}^2P_{1/2},F=3}$, and a Stokes laser driving the transition between $\overline{\ket{3/2^+,{}^2S_{1/2},F=2}}$ and $\ket{5/2^+,{}^2P_{1/2},F=3}$. The latter transition is allowed due to the NHM effect; see the text.
    }
    \label{Fig:STIRAP_Li-likeTh}
\end{figure}

In STIRAP \cite{Bergmann1998_STIRAP, bergmann_perspective_2015}, two coherent laser fields, comprising the pump and the Stokes pulses, couple to a three-level $\Lambda$ system as shown in Fig.\,\ref{Fig:STIRAP_Li-likeTh}(a).
STIRAP is characterized by its robustness against fluctuations of experimental parameters and is independent of spontaneous decay rates and the branching ratio of the upper state.

During the STIRAP process a so-called dark state \cite{Bergmann1998_STIRAP}
\begin{equation} \label{Eq:STIRAP_dark_state}
        \ket{D}=\frac{\Omega_\textrm{S} \left(t \right)}{\sqrt{\Omega_\textrm{S} \left(t \right)^2 +\Omega_\textrm{P} \left(t \right)^2}}\ket{1}- \frac{\Omega_\textrm{P} \left(t \right)}{\sqrt{\Omega_\textrm{S} \left(t \right)^2 +\Omega_\textrm{P} \left(t \right)^2}}\ket{3}
\end{equation}
is formed, with 
$\Omega_\textrm{S/P}$ being the time dependent Rabi frequency of the Stokes/pump pulse.
By adjusting laser parameters such as the delay time $\Delta \tau$ and the pulse energy, the entire population can be transferred from $\ket{1}$ to $\ket{3}$ via the time-dependent dark state without occupying the intermediate state $\ket{2}$.

To give a concrete example, we consider our system initially in the  $\ket{1}=\ket{{}^2S_{1/2},5/2^+,F=3,M_F=3}$ state. This corresponds to fully polarized thorium ions, which can be produced via optical pumping, see also Sec.\,\ref{Subsec:elec_tran}.  Our three-level $\Lambda$ scheme is shown in Fig.\,\ref{Fig:STIRAP_Li-likeTh}. A pump laser couples the initial state $\ket{1}$ and the intermediate state $\ket{2}=\ket{{}^2P_{1/2},5/2^+,F=3,M_F=2}$, while the  Stokes laser couples the state
$\ket{2}$ and the final state $\ket{3}=\overline{\ket{{}^2S_{1/2},3/2^+,F=2,M_F=1}}$, which, due to the NHM, is a superposition state of $\ket{{}^2S_{1/2},3/2^+,F=2,M_F=1}$ and $\ket{{}^2S_{1/2},5/2^+,F=2,M_F=1}$; see Eq.\,\eqref{Eq:NHM}.

The energies and radiative widths of transitions between HFS levels of ${}^2S_{1/2}$ and ${}^2P_{1/2}$ are presented in Table\,\ref{tab:hfs_tran} for Li-like $^{229g}$Th and $^{229m}$Th, respectively. These transitions do not involve change of the nuclear states. Transitions changing both the nuclear and electronic state are presented in Table\,\ref{tab:rad_width_NHM}.
The reduced transition dipole moment of these $E1$ transitions is related to the radiative width, see Eq.\,\eqref{Eq:E1_dipole}.

The Rabi frequency in the ion frame corresponding to the pump laser is given by
\begin{equation} \label{Eq:E1_Rabi_pump}
\begin{split}
    \Omega_\textrm{P}&= \frac{d_\textrm{P}E'}{\hbar}
    =\frac{|\braket{F_gM_{F_g}L\lambda|F_eM_{F_e}}|}{g_{F_e}^{1/2}}\frac{d_rE'}{\hbar}\\
    &=|\braket{F_gM_{F_g}L\lambda|F_eM_{F_e}}|\left(
    \frac{6\pi c^2I_\textrm{P}'\Gamma_{F_e\rightarrow F_g}}{\hbar^2\omega_\textrm{P}'^3}
    \right)^{1/2},
\end{split}
\end{equation}
where $d_\textrm{P}$ is the transition dipole moment amplitude, $\braket{F_gM_{F_g}L\lambda|F_eM_{F_e}}=\braket{331-1|32}$, $\hbar\omega_\textrm{P}'\approx271\,$eV is the transition energy,
$\Gamma_{F_e\rightarrow F_g}=5.0\times10^{-6}\,$eV
is the radiative width of the transition $\ket{{}^2P_{1/2}, F_e}\rightarrow \ket{{}^2S_{1/2}, F_g}$
in Li-like $^{229g}$Th ions, $I_\textrm{P}'(t)=I_{\textrm{P}0}'\exp{[-(4\ln{2}) (t-\tau_\textrm{P}')^2/t_\textrm{p}'^2]}$ is the intensity of the pump-laser pulse, which is assumed to have a Gaussian shape, with $\tau_\textrm{P}'$ denoting the peak position, and $E'=[2I_\textrm{P}'(t)/\epsilon_0 c]^{1/2}$ is the time-dependent amplitude of the electric field in the ion frame with $\epsilon_0$ being the vacuum permittivity.

The reduced transition dipole moment of the transition driven by the Stokes laser, considering the NHM, can be estimated as $d_r$ of the transition between $\ket{2}$ and $\ket{{}^2S_{1/2},5/2^+,F=2,M_F=1}$ multiplied by the NHM coefficient; see Eq.\,\eqref{Eq:red_dip_NHM}.
Consequently, the Rabi frequency corresponding to the Stokes laser can be derived as 
\begin{equation} \label{Eq:E1_Rabi_Stokes}
\begin{split}
    \Omega_\textrm{S}= |b_g|\cdot|\braket{F_gM_{F_g}L\lambda|F_eM_{F_e}}|\left(
    \frac{6\pi c^2I_\textrm{S}'\Gamma_{F_e\rightarrow F_g}}{\hbar^2\omega_\textrm{P}'^3}
    \right)^{1/2},
\end{split}
\end{equation}
where $\braket{F_gM_{F_g}L\lambda|F_eM_{F_e}}=\braket{2111|32}$, $\Gamma_{F_e\rightarrow F_g}=6.2\times10^{-6}\,$eV for Li-like $^{229g}$Th ions,
$I_\textrm{S}'(t)=I_{\textrm{S}0}'\exp{[-(4\ln{2}) (t-\tau_\textrm{S}')^2/t_\textrm{p}'^2]}$ with $\tau_\textrm{S}'=\tau_\textrm{P}'-\Delta\tau'$, where \mbox{$\Delta\tau'\geq0$} is the delay of the pump pulse relative to the Stokes pulse.

The excitation scheme in Fig.\,\ref{Fig:STIRAP_Li-likeTh} could be implemented at the SPS using relativistic Li-like thorium ions with $\gamma=100$ ($\Delta\gamma/\gamma=10^{-4}$). Other parameters of the ion bunches remain those listed in Table\,\ref{tab:ion-beam_facilities}.
Once more we consider  $S_l=S_b$ and $\Gamma_{\textrm{P}}'=\Gamma_D$ (see Sec.\,\ref{Subsec:reso_exc}), which leads to the laser pulse duration $t_\textrm{p}'=67\,$fs.
The durations of the two pulses with similar photon energies are approximately the same.
The laser photon energies are about $\hbar\omega_{\textrm{P}}=1.36\,$eV and $\hbar\omega_{\textrm{S}}=1.31\,$eV.

The population transfer in a Raman scheme can be accomplished by implementing two consecutive $\pi$-pulses (a pump pulse followed by a Stokes pulse in contrast to the STIRAP case) such that
$\int\Omega_\textrm{P}(t)dt=\pi$ and $\int\Omega_\textrm{S}(t)dt=\pi$, requiring $I_{\textrm{P}0}'=2.2\times10^{18}\,$W/m$^2$ and $I_{\textrm{S}0}'=2.8\times10^{22}\,$W/m$^2$.
The latter corresponds to a pulse energy of $E_\textrm{p}=7.8\,$J.
The lifetime of the intermediate state $\ket{2}$ is $\tau_2\approx5.9\times10^{-11}\,$s, much longer than $t_\textrm{p}'$.

STIRAP is generally a more robust approach to population transfer and does not require driving a precise $\pi$-pulse.
To show the population transfer, we solve the master equation in a simplified case neglecting damping as Eq.\,\eqref{Eq:DM_TLS_v2}. The relaxation due to the spontaneous radiative decay of the intermediate state is neglected during the STIRAP process because the lifetime of the intermediate state $\tau_{2}$ is much longer than the pulse duration and delay between the two pulses  ($t_\textrm{p}'/\tau_{2}\approx10^{-3}\ll1$). We are now dealing with a three-level system, with $H_{\textrm{eff}}$ being \cite{Bergmann1998_STIRAP}
\begin{equation}\label{Eq:DM_lambda}
    H_{\textrm{eff}} = -\hbar\begin{pmatrix}
    0   &\frac{\Omega_\textrm{P}(t)}{2}  &0\\
    \frac{\Omega_\textrm{P}(t)}{2} & \Delta_\textrm{P}  &\frac{\Omega_\textrm{S}(t)}{2}\\
    0   &\frac{\Omega_\textrm{S}(t)}{2} &\delta
    \end{pmatrix}\,,
\end{equation}
where $\Delta_\textrm{P(S)}=2\gamma\cdot\omega_\textrm{P(S)}-E_{12(32)}$ is the single photon detuning, $\delta = \Delta_\textrm{P} - \Delta_\textrm{S} $ is the so-called two-photon detuning, $E_{12(32)}$ is the energy difference between the states $\ket{1}(\ket{3})$ and $\ket{2}$.
We again assume searching for the resonance, thus $2\gamma^{\textrm{ave}}\cdot\hbar\omega_\textrm{P(S)}\approx E_{12(32)}$, and consider $\delta\approx0$.

\begin{figure}[!htpb]
    \centering
    \includegraphics[width=\linewidth]{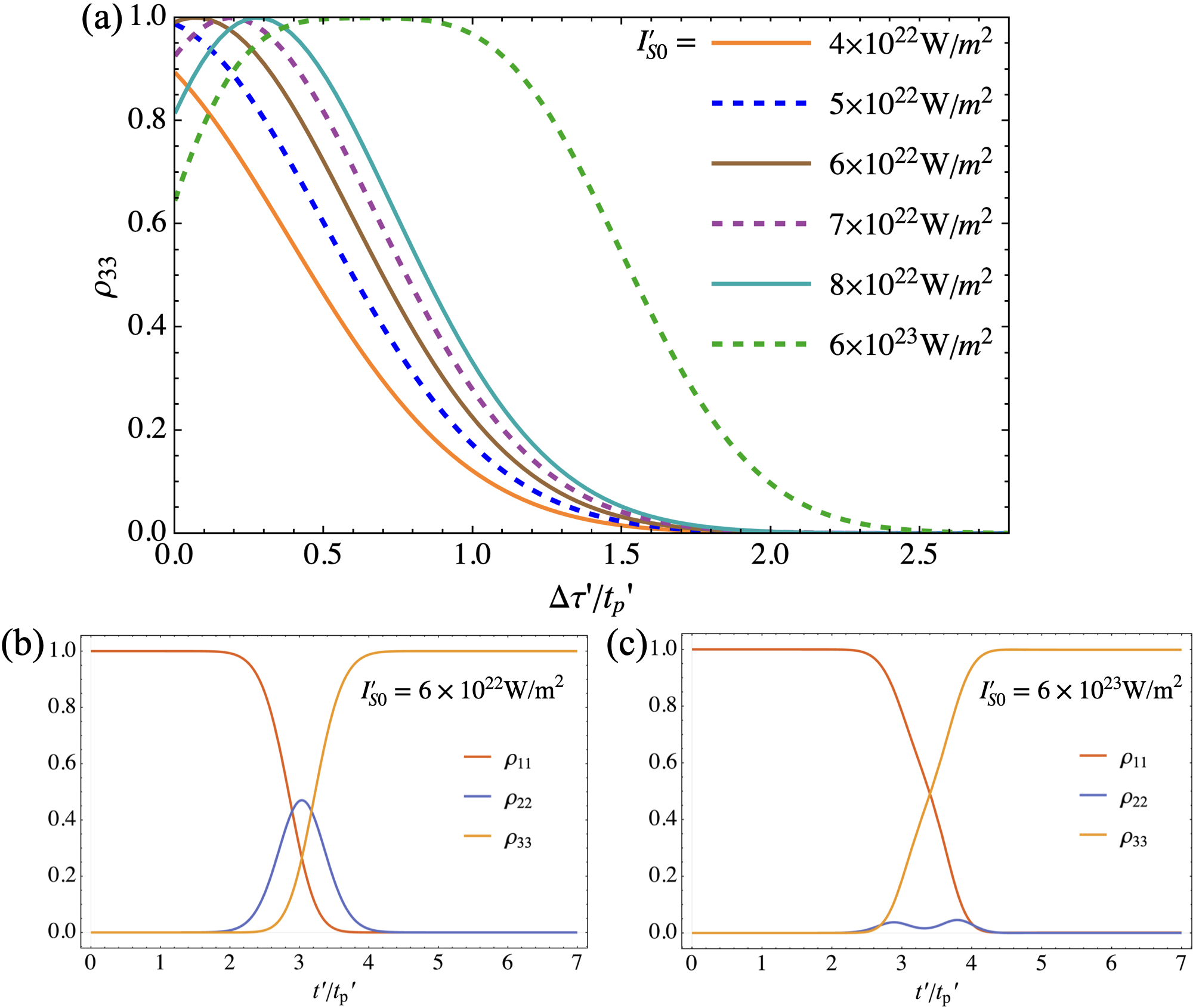}
    \caption{Population transfer into the isomeric state $\ket{3}$ via the STIRAP set-up. (a) Population transfer under different laser-pulse intensities and pulse delays $\Delta\tau'$ (given in the unit of $t_{\textrm{p}}'$). (b) Population evolution under $I_{\textrm{S}0}'=6\times10^{22}\,$W/m$^2$ with delay $\Delta\tau'=0.07\,t_{\textrm{p}}'$ (the minimum $I_{\textrm{S}0}'$ for 100\% population transfer). $\rho_{ii}$ denotes the relative population of the state $\ket{i}$.
    (c) Similar to (b) but for $I_{\textrm{S}0}'=6\times10^{23}\,$W/m$^2$ and $\Delta\tau'=0.76\,t_{\textrm{p}}'$.
    The laser intensities are given for $I_{\textrm{S}0}'$. $I_{\textrm{P}0}'$ is related to $I_{\textrm{S}0}'$ by letting $\Omega_\textrm{P}$ and $\Omega_\textrm{S}$ have the same peak value.
    }
    \label{Fig:STIRAP_Intensity_Delay}
\end{figure}

Through adjusting the laser-pulse intensities and the delay between the two pulses, we identify the lowest $I_{\textrm{S}0}'$ needed for achieving nearly 100\% population transfer as $I_{\textrm{S}0}'\approx6\times10^{22}\,$W/m$^2$ with the delay $\Delta\tau'=0.07\,t_{\textrm{p}}'$; see Fig.\,\ref{Fig:STIRAP_Intensity_Delay}\,(a). The evolution of populations in different states under these optimal parameters is shown in Fig.\,\ref{Fig:STIRAP_Intensity_Delay}\,(b), where the transient population in the intermediate state implies that the transfer process is not ideally adiabatic, which should be limited by the laser-pulse intensity.
For instance, if $I_{\textrm{S}0}'$ is increased by an order of magnitude together with a suitable delay between the two pulses, the maximal transient population of the intermediate state would be smaller than five percent, see Fig.\,\ref{Fig:STIRAP_Intensity_Delay}\,(c).
An empirical criterion based on the adiabaticity condition \cite{bergmann_perspective_2015,Vitanov_2017} gives the order of magnitude required for the peak Rabi frequency amplitude,
$\Omega_0^2 \geq 200\ln{2}/t_{\textrm{p}}'^2$,
for coherent pulses, leading to $\widetilde{I_\textrm{S0}}'\approx9\times10^{23}\,$W/m$^2$ and
$\widetilde{I_\textrm{P0}}'\approx7\times10^{19}\,$W/m$^2$.

We note that in Fig.\,\ref{Fig:STIRAP_Intensity_Delay}, $I_{\textrm{P}0}'$ is related to $I_{\textrm{S}0}'$ by letting the peak value of $\Omega_\textrm{P}$ equal to that of $\Omega_\textrm{S}$. The minimum $I_{\textrm{S}0}'\approx6\times10^{22}\,$W/m$^2$ corresponds to $I_{\textrm{P}0}'\approx4.8\times10^{18}\,$W/m$^2$.
A larger $I_{\textrm{P}0}'$, for instance, increased by a factor of five, would actually reduce the population transfer, and would require a larger $I_{\textrm{S}0}'$ instead of relieving the requirement to achieve again the 100\% population transfer. Reducing $I_{\textrm{P}0}'$ has a similar effect.

The population transfer displayed in Fig.\,\ref{Fig:STIRAP_Li-likeTh} is for ions corresponding to $\Delta_\textrm{P}\approx\Delta_\textrm{S}\approx0$.
The ion-energy spread can be further considered using Eq.\,\eqref{Eq:ave_popu_tran}, resulting in a relatively small correction, since we have adjusted the pulse duration to match the laser-pulse spectrum and the photon-absorption spectrum considering the ion-energy spread.
For instance, the 100\% population transfer for the scenario shown in
Fig.\,\ref{Fig:STIRAP_Intensity_Delay}\,(b) corresponds to an average population transfer of $\rho_0=85\%$ when considering the ion energy spread.
The laser-beam parameters and corresponding excitation rates are presented in Table\,\ref{tab:elec_STIRAP}.

\begin{table*}[htpb]
    \centering
    \begin{tabular*}{\linewidth}{@{\extracolsep{\fill}} lcc ccccc}
    \hline 
    \hline
    $\gamma$    &$\hbar\omega_\textrm{S}\,$(eV)  &$\hbar\omega_\textrm{P}\,$(eV)   &$I_\textrm{S0}\,$(W/m$^2$)  &$I_\textrm{P0}\,$(W/m$^2$) &$t_\textrm{p}\,$(ps)    &$\rho_0$    &rate\,(s$^{-1}$) \\
    \hline \\[-0.2cm]
    100  &1.31   &1.36  &$1.5\times10^{18}$  &$1.2\times10^{14}$   &13    &0.85   &$1.3\times10^{14}$\\
    \hline
    \hline
    \end{tabular*}
    \caption{
    Isomer excitation via STIRAP setup at the SPS using an intermediate electronic excited state in Li-like $^{229}$Th ions. The laser-beam parameters, including the photon energy $\hbar\omega_\textrm{P/S}$, peak intensity $I_{\textrm{S/P0}}$ and pulse duration $t_\textrm{p}$,  are given in the lab frame,
    related to the ion-frame parameters according to Eqs.\,\eqref{Eq:para_frame}. $\rho_0$ is the isomer excitation fraction considering the ion-energy spread.
    The excitation scheme and the population transfer are displayed in Figs.\,\ref{Fig:STIRAP_Li-likeTh} and \ref{Fig:STIRAP_Intensity_Delay}\,(b), respectively.
    Note that for real STIRAP with almost no transient population of the intermediate state, one order of magnitude higher intensity would be needed; see Fig.\,\ref{Fig:STIRAP_Intensity_Delay}\,(c).
    The average isomer excitation rate in the last column is estimated as $\rho_0 N_{\textrm{ion}} N_b f_b$.}
    \label{tab:elec_STIRAP}
\end{table*}

Due to the NHM, one can also excite the transitions with energies of about 279\,eV, exciting both the nuclear and electronic states (see Fig.\,\ref{Fig:Li-like_Th}).
Therefore, in the STIRAP scheme shown in Fig.\,\ref{Fig:STIRAP_Li-likeTh}, we can also use the state $\overline{\ket{{}^2P_{1/2},3/2^+,F=2}}$ as the intermediate state.

If one does not start with fully polarized ions, population in $\overline{\ket{{}^2S_{1/2},5/2^+,F=2}}$ level can be transferred to the isomeric state, $\overline{\ket{{}^2S_{1/2},3/2^+,F=2}}$, via STIRAP with any of the four HFS levels of ${}^2P_{1/2}$
being the intermediate state.

\subsubsection{
Isomer excitation via the 279 eV level using a single laser
}
\label{subsubsec:exc_279eV}

Besides STIRAP, one can also excite the nuclear isomeric state using only a single laser to drive the 279\,eV transition (see Fig.\,\ref{Fig:Li-like_Th}), which involves simultaneous excitation of both the nuclear and electronic states, enabled by the NHM effect.
The required laser-pulse intensity for implementing a $\pi-$pulse (between two HFS levels) can be derived using Eq.\,\eqref{Eq:Intensity_pi} and the radiative widths listed in Table\,\ref{tab:rad_width_NHM}.

After driving the transition, thorium ions will first quickly decay ($T_{\xi_eJ_e}=5.9\times10^{-11}\,$s in the ion frame)  from the electronic excited state to the electronic ground state but remain in the nuclear isomeric state which has a much longer lifetime (about 1s in the ion frame).
Detecting the spontaneously emitted photons from the 271\,eV electronic transition after exciting the 279\,eV transition would unambiguously demonstrate isomer excitation. This would also serve as an experimental demonstration of the expected NHM effect in $^{229}$Th ions.
We note that, after excitation, the branching ratio of direct decay into the state with both the electron and nucleus in the ground state is negligible, on the order of $10^{-5}$; see Table\,\ref{tab:rad_width_NHM}.

While an excitation scheme using a single laser beam is, in principle, simpler to implement, STIRAP may still be beneficial for certain experiments. Indeed, the method is tolerant to single-photon detuning and provides a robust and complete transfer to the isomeric state. Additionally, since STIRAP is a coherent process, all ions will experience the same photon recoil during the excitation process, while spontaneous-emission step leads to a random ``kick.'' The latter may become important when deep laser cooling is implemented in stored highly relativistic HCI \cite{Winters-2015a,Eidam2018LaserCooling}.

\subsection{Excitation via the 29~keV level}
\label{Subsec:exc_29keV}

While direct resonant excitation of the isomeric state can, in principle, be implemented at many accelerators,
laser ($\sim$eV) excitation via the 29\,keV level requires a relatively large Lorentz factor, which can be achieved at a smaller number of accelerators, e.g., the LHC. However, in contrast to the large uncertainty in the energy of the isomeric state, the energy of the second nuclear excited state has been measured precisely as
$E_{\textrm{2nd}}=29,189.93(7)$\,eV \cite{Masuda2019ThXray} with a relative uncertainty about $2.4\times10^{-6}$, smaller than $\Delta\gamma/\gamma$ typically achievable at storage rings.
Therefore, the resonance can be found more easily compared to the direct excitation scenario.
The precise knowledge of the energy of the 29\,keV level would also allow one to determine the central value of $\gamma$, $\gamma^\textrm{ave}$, with a similar relative precision via detection of secondary photons from radiative decay of this state; see Sec.\,\ref{Subsec:Fur_exc}, thereby removing one of the major systematic uncertainties in storage-ring based laser spectroscopy experiments\,\cite{Ullmann-2017,Lochmann-2014}.
Some other key properties of this state are shown in Fig.\,\ref{fig:229Th}.

\subsubsection{Incoherent population transfer}
\label{subsubsec:inco_exc_29keV}

X-ray pumping of the isomer state via the 29\,keV level has been demonstrated in Ref.\,\cite{Masuda2019ThXray}, where $^{229}$Th nuclei in a solid target were excited using 29\,keV synchrotron radiation.
Here, we discuss laser excitation of the isomeric state using relativistic $^{229}$Th ions.
$^{229}$Th ions with
$\gamma\approx2950$
($\Delta\gamma/\gamma\approx10^{-4}$) can be obtained
at the LHC; see Table\,\ref{tab:ion-beam_facilities}.
Thereby a laser with photon energy
$\hbar\omega_{12}\approx4.95\,$eV
can be used for resonant excitation of this transition with nearly head-on collisions between laser photons and the ion bunch.

Suppose that we employ a pulsed laser with an energy of 10\,$\mu$J per pulse and a repetition rate of $f_l=N_b f_b\approx13.8\,$MHz ($N_b=1232$) synchronized to the ion bunch, and a cavity further enhancing the laser power by a factor of 10$^5$ (leading to $E_{\textrm{p}}=1\,$J).
We consider again $S_l=S_b$ and $\Gamma_{\textrm{P}}'=\Gamma_D$ (see Sec.\,\ref{Subsec:reso_exc}),
which leads to a pulse duration $t_\textrm{p}\approx3.7\,$ps and peak intensity $I_0\approx3.2\times10^{20}\,$W/m$^2$.

When exciting the $M1$ transition from the nuclear ground state to the 29\,keV state, the system can be approximated as a two-level system since the radiative decay of the excited state is negligible due to its long lifetime compared to the pulse duration.
The population evolution induced by individual laser pulses can be derived by solving Eq.\,\eqref{Eq:DM_TLS_v2}.
The Rabi frequency of the transition can be derived from Eq.\,\eqref{Eq:Rabi_red} with $\Gamma_\textrm{rad}$ replaced by the partial width $\Gamma_{\gamma}^\textrm{cr}\approx 1.70\,$neV in this case.
After the ion bunch is excited by a laser pulse, it circulates in the storage ring before interacting with the next laser pulse.  One needs to consider the spontaneous radiative decay of the 29\,keV level during this period, which could end up in both the nuclear ground and isomeric states with the relative radiative branching ratios being about 9\% and 91\%, respectively.
Only radiative branching ratios are considered here because IC channels are energetically forbidden in highly charged ${}^{229}$Th ions (only $2p_{3/2}$ or higher orbitals have ionization energies smaller than 29\,keV \cite{NIST_ASD}).

\begin{table}[htpb]
    \centering
    \begin{tabular*}{\linewidth}{@{\extracolsep{\fill}} lcccc c}
    \hline 
    \hline
    $\gamma$    &$t_\textrm{p}\,$(ps)    &$I_0\,$(W/m$^2$)  &$\rho_{ee}(\Delta=0)$   &$\rho_0$   &$t_\textrm{ex}\,$(s) \\
    \hline \\[-0.2cm]
    2950   &3.7   &$3.2\times10^{20}$  &$6.0\times10^{-4}$   &$4.2\times10^{-4}$   &0.56\\
    \hline
    \hline
    \end{tabular*}
    \caption{
    Repeated pumping of $^{229m}$Th isomers via the 29\,keV level at the LHC. $\gamma$, $t_\textrm{p}$ and $I_0$ are some of the considered parameters of the ion bunch and laser beam. $\rho_{ee}(\Delta=0)$ and $\rho_0$ are the population-transfer fraction of the ion group resonant with the laser pulse and the transfer fraction averaged over all ions, respectively, in the first round.
    $t_\textrm{ex}$ is the time needed to achieve $99\%$ of the maximal population transfer, $\rho_{\textrm{max}}\approx1$, corresponding to $\approx10^8$ ions per bunch. If one considers implementing a $\pi-$pulse,  the required intensity is $I_{0\pi}\approx2100 I_0$.
    }
    \label{tab:repeat_pump_lhc}
\end{table}

\begin{figure}[!htpb]\centering
    \includegraphics[width=0.9\linewidth]{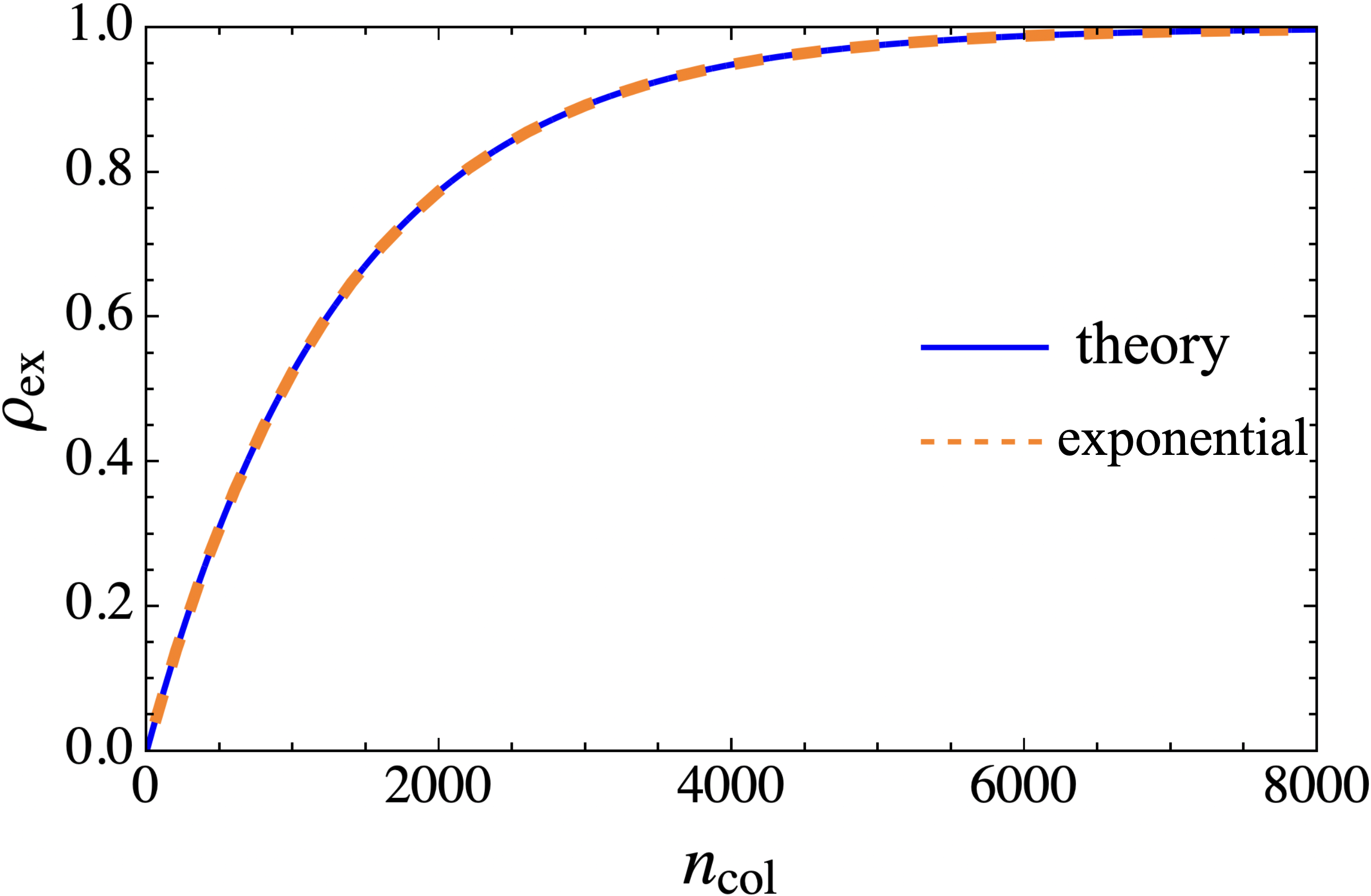}
    \caption{
    Repeated pumping at the LHC. The blue solid curve is derived from the theoretical calculation described in the text, which also agrees with Eq.\,\eqref{Eq:exp_behavior} shown as the orange dashed curve.
    Nearly full population transfer is reached after $n_\textrm{col}\approx6300$ collisions ($t_\textrm{ex}\approx0.56\,$s).
    }
    \label{fig:Rep_pump_LHC}
\end{figure}

The parameters and results of repeated isomer excitation via the 29\,keV level at LHC are presented in Table\,\ref{tab:repeat_pump_lhc} and Fig.\,\ref{fig:Rep_pump_LHC}.
The increase of $\rho_{\textrm{ex}}$ can be approximated using Eq.\,\eqref{Eq:exp_behavior}, but with $\rho_\textrm{max}\approx1$ and $\rho_0$ substituted by $\rho_0/f_\textrm{dec}$, because we are now considering a three-level system.
$f_\textrm{dec} = 1-\exp{[-(1/f_b)/(\gamma\cdot T_{1/2}^{\textrm{rad}}/\ln{2})]} = 0.57$
is the decay fraction of ions in the 29\,keV level during one circulation in the storage ring and $\gamma\cdot T_{1/2}^{\textrm{rad}}/\ln{2}$ is the lab-frame radiative lifetime of the 29\,keV level.

\subsubsection{STIRAP via the 29~keV state}
\label{subsubsec:STIRAP_29keV}

We now consider exciting $^{229m}$Th isomers via STIRAP, which is similar to the scheme in Sec.\,\ref{subsubsec:STIRAP_Li-like}, but uses an intermediate nuclear state at 29 keV instead of an electronic excited state and does not rely on the NMH.
This scenario at the GF was proposed in Ref.\,\cite{Budker2021GF_nucl} and investigated theoretically in more detail in Ref.\,\cite{Kirschbaum2022}. The corresponding nuclear transition energies are $E_{12}=E_\textrm{2nd}\approx29.190\,$keV and $E_{32}=E_\textrm{2nd}-E_{\textrm{iso}}\approx29.182\,$keV. $\ket{1}$, $\ket{2}$ and $\ket{3}$ denote the nuclear ground, second excited and isomeric state, respectively.
Considering  $\gamma\approx2950$ at the LHC, the required photon energies for the pump and Stokes pulses are $\hbar\omega_\textrm{P}\approx\hbar\omega_\textrm{S}\approx 4.95\,$eV. Following Sec.\,\ref{Subsec:reso_exc}, we set $S_l=S_b$ and $\Gamma_\textrm{p}'=\Gamma_D$, leading to FWHM duration $t_{\textrm{p}}\approx\SI{3.7}{\pico \second}$ for both Gaussian shaped pulses.

The coherent population-transfer dynamics can be evaluated using the density matrix formalism following a procedure similar to Sec.\,\ref{subsubsec:STIRAP_Li-like}, solving Eq.\,\eqref{Eq:DM_TLS_v2} with the Hamiltonian given in Eq.\,\eqref{Eq:DM_lambda}. We neglect the radiative decay of the 29\,keV state, since the radiative lifetime is much longer than the pulse duration and the delay between the Stokes and pump pulses.
The time-dependent Stokes and pump Rabi frequencies are given by Eq.\,\eqref{Eq:Rabi_red}, with the corresponding partial radiative widths  being $\Gamma_{\gamma}^{\textrm{in}}\approx 16.6\,$neV and $\Gamma_{\gamma}^{\textrm{cr}}\approx 1.70\,$neV, respectively (see the beginning of Sec.\,\ref{Sec:iso_exc}).

The empirical adiabaticity criterion \cite{bergmann_perspective_2015,Vitanov_2017} mentioned in Sec.\,\ref{subsubsec:STIRAP_Li-like} provides the peak intensities required for the pump and Stokes pulses for robust STIRAP,
$\widetilde{I_\textrm{P0}'}\approx 7.4\times10^{32}$\,W/m$^2$ 
and  $\widetilde{I_\textrm{S0}'}\approx 7.6\times10^{31}$\,W/m$^2$, corresponding to
$\widetilde{I_\textrm{P0}}\approx 2.1\times10^{25}$\,W/m$^2$ (pulse energy $E_\textrm{p}\approx6.6\times10^4\,$J)
and  $\widetilde{I_\textrm{S0}}\approx 2.2\times10^{24}$\,W/m$^2$
in the lab frame. These intensities are orders of magnitude higher than the $I_0$ used for repeated incoherent pumping in Sec.\,\ref{subsubsec:inco_exc_29keV}. However, the high laser-pulse repetition rate required for accumulating excitation in repeated pumping becomes unnecessary here.

\begin{figure}[!htpb]
    \centering
    \includegraphics[width=0.9\linewidth]{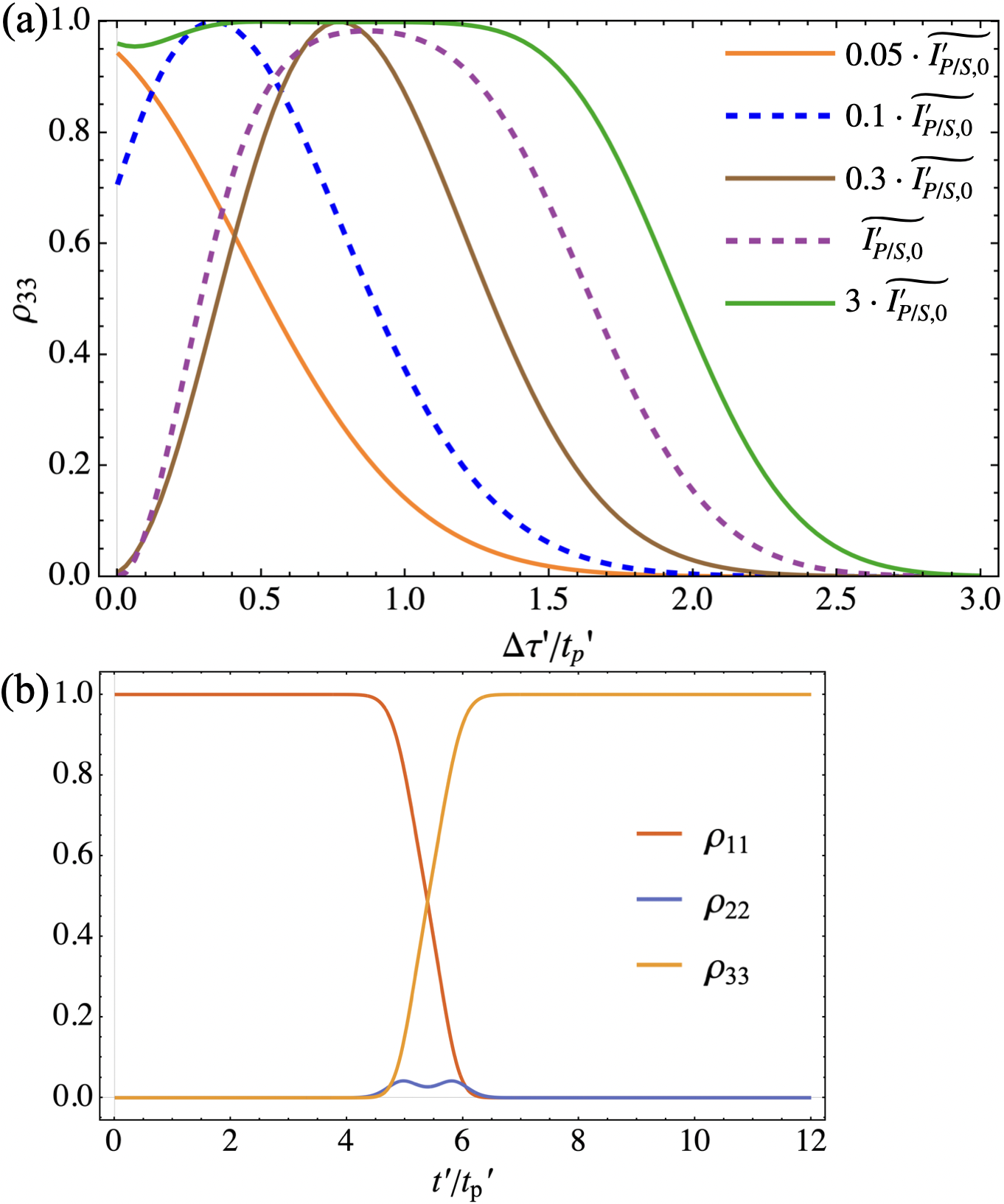}
    \caption{
    (a) Population transfer into the isomeric state $\ket{3}$ via the STIRAP set-up under different laser-pulse intensities and pulse delays.
    (b)
    Population evolution under $0.6\cdot \widetilde{I_{\textrm{P/S,0}}'}$ and $\Delta\tau'=0.78\,t_{\textrm{p}}'$.
    }
    \label{Fig:STIRAP_lhc}
\end{figure}

Figure\,\ref{Fig:STIRAP_lhc}\,(a) shows the population transfer to state $\ket{3}$ after a Stokes and pump pulse sequence, under different laser-pulse intensities and different pulse delays $\Delta\tau$.
Nearly full population transfer can be reached for a large range of $\Delta\tau$ using the peak intensities given above.
For lower intensities, nearly full population transfer can still be achieved but in a narrower range of $\Delta\tau$. 
Figure\,\ref{Fig:STIRAP_lhc}\,(b) shows an example where full population transfer into state $\ket{3}$ with almost no transient population of state $\ket{2}$ is achieved for peak intensity values $0.6\cdot\widetilde{I_\textrm{P/S,0}'}$ with $\Delta\tau'=0.78\,t_{\textrm{p}}'$.
Similar to the discussion in Sec.\,\ref{subsubsec:STIRAP_Li-like}, 
the population transfer shown in Fig.\,\ref{Fig:STIRAP_lhc} is calculated for ions that satisfy the condition $\Delta_\textrm{P}\approx\Delta_\textrm{S}\approx0$. Ion-energy spread results in only a relatively small correction.

A more detailed discussion of the available ion bunch and laser beam parameters for this STIRAP scenario is presented in Ref.\,\cite{Kirschbaum2022}. Given that the 29\,keV state has a well-known energy and a favorable branching ratio to the isomeric state, repeated incoherent pumping could be more favorable for near-term experimental implementation. Its simplicity (lower laser intensity and only one laser frequency) should be weighed against more stringent requirements for laser-frequency tuning as opposed to robust STIRAP tolerant to laser detuning.

\subsection{Isomer excitation with x-ray lasers}
\label{subsec:exc_x_ray}

Some schemes discussed above (those in Sec\,\ref{Subsec:exc_271eV} and \ref{Subsec:exc_29keV}) could also be realized using x-ray pulses combined with moderately accelerated or even static ${}^{229}$Th ions.

The photon energy needed for the schemes discussed in Sec\,\ref{Subsec:exc_271eV} is about $270\,$eV, well within reach of many x-ray laser facilities.
Excitation via STIRAP can be done by use of a two-color x-ray free electron laser (FEL). These devices are capable of generating intense x-ray pulses by guiding relativistic bunches of electrons through a periodic array of dipole magnets known as an undulator~\cite{mcneilXrayFreeelectronLasers2010}. State-of-the-art FELs allow for the generation of two pulses of different energies with tunable time delays~\cite{pratWidelyTunableTwocolor2022}, which potentially can be used to perform STIRAP operations. Such experiments have been proposed for purely nuclear transitions~\cite{liaoNuclearCoherentPopulation2011, mansourzadeh-ashkaniSuperpositionNuclearStates2022} and for electronic ones~\cite{piconStimulatedRamanAdiabatic2015}.
One can also use just a single x-ray laser to excite the 279\,eV transition to transfer population into the isomeric state, following the scheme discussed in Sec.\,\ref{subsubsec:exc_279eV}.

The highly charged ions could be produced and trapped using an electron beam ion trap (EBIT) installed as an end station of an FEL. The EBIT operates by means of a magnetically focused electron beam which ionizes neutral particles which are introduced as, for example, a tenuous atomic beam~\cite{levineElectronBeamIon1988}. Ions remain trapped in the negative space-charge potential of the electron beam where they are sequentially ionized further until the ionization energy is beyond the electron beam energy. Production of HCI up to H-like uranium was demonstrated\,\cite{marrsProductionTrappingHydrogenlike1994}; production of Li-like thorium is possible with transportable devices such as the FLASH-EBIT, which has already been used for one-photon spectroscopy at FELs\,\cite{epp2007rontgen, bernittUnexpectedlyLowOscillator2012}. 
We stress, however, that due to the need to produce $^{229}$Th in a nuclear reaction,
the number of ions will inevitably be tiny and also unavoidable contaminants needs to be considered.
Subsequent to the STIRAP or single-laser excitation, the state of the thorium nucleus can be determined in various ways. For example, this can be done by extracting the ions and neutralizing them using an electrode or ultrathin carbon foil to open the internal conversion (IC) channel which results in detectable IC electrons in case the isomer would have been excited~\cite{Bilous2020EB_35+}.

We note that isomer excitation via the 29\,keV level using x-ray lasers is discussed in Ref.\,\cite{Kirschbaum2022}.

\section{Detection of the thorium isomer}
\label{Sec:Detect}

Besides excitation of $^{229m}$Th isomers discussed above, accelerators with relativistic HCI also offer unique opportunities for detection of those isomers and precision measurements of the isomeric-state energy.

We now present three different schemes for isomer detection, relying on (i)
the radiative decay of $^{229m}$Th, (ii) further excitation of the isomeric state to the 29\,keV level followed by radiative decay, and (iii) laser spectroscopy of electronic transitions.
In contrast to the first two approaches, the third one requires using partially stripped ions (PSI) and electronic transitions instead of only nuclear transitions.
Besides the demonstration of isomer excitation, we also discuss the precision that can be achieved in each scheme in  determining the energy of the isomeric state.

\subsection{Radiative decay of the isomeric state}
\label{Subsec:Rad_decay}
Until recently \cite{Kraemer2022_obs_rad_decay}, there was no successful and unambiguous detection of photons from the radiative decay of $^{229m}$Th; see, for example, experiments in Refs.\,\cite{Jeet2015DirectExc,Yamaguchi2015DirectExc,Stellmer2018OpticalExc}. This could be due to unexpected non-radiative decay processes in the crystal/solid-state environment of these experiments \cite{Peik2021}.
In contrast, in HCI
radiative decay should be the predominant decay channel of $^{229m}$Th.

Photons from radiative decay of the isomeric state
will not be emitted at a fixed location in the storage ring but rather at arbitrary locations along the ring perimeter during the circulation of the ion bunch. This is due to the long radiative half-life of $^{229m}$Th, even when there is lifetime reduction due to the NHM effect.

The average photon-detection rate can be estimated as
\begin{equation}
\label{Eq:det_rad}
    p_\text{d}=N_{\text{iso}}N_{\text{b}}f_{\text{b}}\left[\exp{\left(-\frac{t_1}{\tau}\right)}-\exp{\left(-\frac{t_2}{\tau}\right)}\right]\approx N_{\text{iso}}N_{\text{b}}f_{\text{b}}\frac{\Delta t}{\tau}\,,
\end{equation}
where $N_{\text{iso}}$ is the number of isomers per bunch, $\tau=\gamma T_{1/2}^\textrm{rad}/\ln{2}$ is the radiative lifetime of the isomeric state in the lab frame, and $\Delta t=t_2-t_1$ is the time duration related to the photon collection during each ion-bunch circulation with $t_1$ and $t_2$ being the starting and ending times, respectively.
We assume that during each circulation, photons emitted during the ions travelling in, say, $1\,$m can be detected, i.e., $c\Delta t=1\,\textrm{m}$. This photon-collection length refers to the lab frame and can be made significantly longer with the help of dedicated mirror systems, see, for example, Ref.\,\cite{Sanchez-2017}.
Here, the time delay $t_1$ ($t_1\ll\tau$) is introduced to help filter out photons from the pulsed excitation laser and to collect photons only from the radiative decay of $^{229m}$Th.
Since the lifetime of $^{229m}$Th is long compared to the laser pulse duration, background associated with the excitation laser could be entirely eliminated.

We note that in the ion frame, photons from spontaneous emission (referred to as secondary photons below) are emitted in a broad angular distribution.
As seen in lab frame, however, due to the relativistic effect, the photons are mainly emitted within a small angle ($\approx1/\gamma$) relative to the ion-propagation direction.
Photons re-emitted along the
ions-propagation direction are boosted in energy by another factor of $\approx2\gamma$, so the maximal energy of the secondary photons is about $2\gamma\cdot\hbar\omega'$; see, for instance, Refs.\,\cite{Budker2020_AdP_GF,Budker2021GF_nucl} for more details.

For an estimate, we assume that almost all ions in a bunch are excited to the isomeric state, i.e., $N_{\textrm{iso}}\approx N_{\textrm{ion}}$ is the number of isomers per ion bunch.
We obtain $p_\textrm{d}\approx3.9\,\textrm{s}^{-1}$, $2.0$\,s$^{-1}$ and 0.21\,s$^{-1}$ at the HESR ($\gamma=6$), SPS ($\gamma=35$) and LHC ($\gamma=2950$), respectively, for bare nuclei ($T_{1/2}^\textrm{rad}\approx5.2\times10^3\,$s).
The low counting rates are mainly due to the large $\tau$, and
can be to some degree improved by using ion bunches with a smaller $\gamma$, leading to shorter lifetime in the lab frame.
$\gamma$ could be reduced to around 2, 10 and 200 at the HESR, SPS and LHC, respectively. Then, we get $p_\textrm{d}\sim10\,$s$^{-1}$ at these accelerators. As mentioned above, higher counting rates could also be obtained through increasing the photon-collection length.

For H-like and Li-like thorium ions with the NHM, $\tau$ could be reduced by a factor of about $10^5$ and $10^4$, respectively, compared to bare nuclei, leading to increase of the photon detection rate by the same factor.
This can also serve as an experimental evidence of the NHM effect, which has not been measured in atoms or atomic ions as yet \cite{Shabaev2021NHM}.

Detection of secondary photons from the radiative decay of the isomeric state directly demonstrates excitation of $^{229m}$Th,
which means the resonance condition for excitation, $E_\textrm{iso}=2\gamma\hbar\omega$,
is fulfilled (($\hbar\omega$ is the laser photon energy)).
Therefore, the isomeric-state energy can be obtained, with a relative precision of $\lesssim\Delta\gamma/\gamma$. This precision corresponds to the uncertainty in $\gamma$ and depends on how well the laser-pulse spectrum 
matches the photon-absorption spectrum considering the ion-energy spread.
After improving the match
(see Sec.\,\ref{Sec:search_res}), $E_\textrm{iso}$ could be determined more precisely (with a relative precision similar to that of $\gamma^{\textrm{ave}}$) through measuring $\gamma^{\textrm{ave}}$.

$\gamma^{\textrm{ave}}$ can be determined via measuring the energy of secondary photons in the lab frame, $E_{121}=2\gamma E_\textrm{iso}$.
From $E_{121}=2\gamma E_\textrm{iso}=4\gamma^2\hbar\omega$,
we derive $\gamma=(E_\textrm{121}/4\hbar\omega)^{1/2}$ and $E_\textrm{iso}=2\gamma\hbar\omega=
(E_{121}\hbar\omega)^{1/2}$; see also Refs.\,\cite{Wojtsekhowski2022local,Budker2021GF_nucl}. $\gamma^{\textrm{ave}}$ can be measured with a relative precision of $1/2(p_\textrm{d} t_\textrm{det})^{1/2}$ with $t_\textrm{det}$ being the measurement time. In the case of bare nuclei, because of the low detection rate, an 1-hour measurement produces $\gamma^{\textrm{ave}}$ with a relative precision of about $3\times10^{-3}$, which gives no improvement compared to $\Delta\gamma/\gamma$.
For H-like thorium ions, an 1-hour measurement produces $\gamma^{\textrm{ave}}$ with a relative precision of $\lesssim10^{-5}$.

\subsection{Further excitation of the isomeric state}
\label{Subsec:Fur_exc}
Since the long radiative half-life of the isomeric state reduces detection rates of secondary photons, it could be more efficient to further excite $^{229m}$Th from the isomeric state to the second nuclear excited state at 29\,keV which has a significantly shorter half-life, $T_{1/2}^\textrm{rad}\approx25\,$ns, although a large $\gamma$ is needed for such excitation.

For the isomer-production process discussed in Sec.\,\ref{subsubsec:inco_exc_29keV},
when exciting the transition from the ground state to the 29\,keV state, there are initially (before a significant fraction of the nuclei are optically pumped)
$N_0\approx N_\textrm{ion}\cdot\rho_0/\textrm{BR}_{\gamma}^{\textrm{in}}/f_\textrm{dec}\approx 8.1\times10^4$
excitations into the 29\,keV state induced by a single laser pulse.
The corresponding average photon-detection rate is
$p_\textrm{d} \approx N_0N_bf_b\Delta t/\tau\approx3.5\times10^7\,$s$^{-1}$
[see Eq.\,\eqref{Eq:det_rad}]. We again assumed a 1-m collection length and a slight delay in detection $t_1\ll\tau$ (here $\tau$ is for the 29\,keV state).

After almost all ions are excited to the isomeric state, which can be known from the decrease of the detection rate of photons from the radiative decay of the 29\,keV level, we can decrease $\gamma$ or the laser-photon energy
to change the ion-frame laser-photon energy by about $E_\textrm{iso}$,
and observe increased photon collection rates due to excitation of the isomeric state.
Excitations of the isomeric state and the ground state can be distinguished since the transition widths are both around 29.2\,keV$\times\Delta\gamma/\gamma\approx2.9\,$eV in the ion frame, while the energy difference between the isomeric and ground state is about 8.2\,eV.
We note that this probing method could also be applied to isomers produced at the PS or SPS, since ion bunches at those facilities can be transferred to the LHC and further accelerated for excitation of the 29\,keV transition.

Besides probing isomer production, a precise determination of the isomeric-state energy can be realized in this process via detecting secondary photons from radiative decay of the 29\,keV level, utilizing the recent precise measurement of the energy of the 29\,keV state \cite{Masuda2019ThXray}.

Following a procedure similar to the one in Sec.\,\ref{subsubsec:inco_exc_29keV},
we obtain excitation events from the isomeric state to the 29\,keV state as
$N_0\approx8.0\times10^5$
induced by one laser pulse for $N_\textrm{iso}\approx10^8$ isomers per bunch.
The average photon-collection rate is 
$p_\textrm{d}\approx3.5\times10^8\,$s$^{-1}$.
Since nearly 9\% of the ions excited to the 29\,keV level will decay into the ground state, the number of isomers per ion bunch as well as the photon-collection rate will decrease within a second. To keep the high photon-collection rate, a second laser can be employed to excite ions from the nuclear ground state to the isomeric state via the 29\,keV level.

We assume two lasers with photon energy $\hbar\omega_{12}$ and $\hbar\omega_{32}$ in the lab frame are
tuned to be resonant with the transitions from the nuclear ground and isomeric state to the 29\,keV level, respectively.
Here, as in Sec\,\ref{subsubsec:STIRAP_29keV}, states $\ket{1}$, $\ket{2}$ and $\ket{3}$ represent the nuclear ground state, 29\,keV state and isomeric state, respectively.
$E_\textrm{iso}$ can be obtained from
\begin{equation}
\label{Eq:mea_29keV}
    E_\textrm{iso}=2\gamma (\hbar\omega_{12}-\hbar\omega_{32})=\frac{E_{\textrm{2nd}}}{\hbar\omega_{12}}(\hbar\omega_{12}-\hbar\omega_{32})\,.
\end{equation}
Here, $2\gamma \hbar\omega_{12}=E_{\textrm{2nd}}$ and $2\gamma \hbar\omega_{32}=E_{\textrm{2nd}}-E_\textrm{iso}$ are the resonance conditions.
After matching the laser-pulse spectrum in the ion frame and the photon-absorption spectrum considering the ion-energy spread (see Sec.\,\ref{Sec:search_res}),
$\gamma=E_{\textrm{2nd}}/2\hbar\omega_{12}$ can be determined with a relative precision 
similar to that of $E_{\textrm{2nd}}$, about $2.4\times10^{-6}$ as measured in Ref.\,\cite{Masuda2019ThXray}. $E_\textrm{iso}=2\gamma (\hbar\omega_{12}-\hbar\omega_{32})$ could be determined with almost the same precision.
We have assumed that $\hbar\omega_{12(32)}^\textrm{ave}$, i.e., the central values of photon energies of the two lasers, can be known with a better precision.
We note that, this scheme, exploiting the recent measurement of $E_{\textrm{2nd}}$, does not require precise knowledge of energies of secondary photons, thus it might be easier to implement experimentally.

A higher precision of $E_{\textrm{iso}}$ could be derived via measuring energies of secondary photons. Secondary photons from the radiative decay of the 29\,keV level into the nuclear ground state and the isomeric state have maximal energies of
$E_{\textrm{321}}=2\gamma E_{\textrm{2nd}}$ and $E_{\textrm{323}}=2\gamma (E_{\textrm{2nd}}-E_{\textrm{iso}})$, respectively, in the lab frame.
Detecting those high-energy photons could be realized using the photon detector described in Ref.\,\cite{Wojtsekhowski2022local}.
With a measurement time of $t_\textrm{det}=1\,$h,
$E_{323}$ can be determined with a relative uncertainty
$\Delta E_{323}/E_{323}\approx
1/(p_\textrm{d}\cdot t_\textrm{det}\cdot\textrm{BR}_\gamma^{\textrm{in}})^{1/2}\approx9.4\times10^{-7}$.
$\gamma=(E_{323}/4\hbar\omega_{32})^{1/2}$
can be determined with a relative precision of $4.7\times10^{-7}$, allowing for measuring $E_\textrm{iso}$ with almost the same precision.
$E_{\textrm{2nd}}$ can also be measured with a similar precision in this process due to
$E_{\textrm{2nd}}= 2\gamma\hbar\omega_{12}$,
which would be better than the precision obtained in Ref.\,\cite{Masuda2019ThXray}.

\subsection{Nuclear spectroscopy via electronic transitions}
\label{Subsec:elec_tran}

Population transfer into the isomeric state can also be probed via laser spectroscopy of electronic transitions,
see, for example, Ref.\,\cite{Peik2003_spectroscopy}.
The electronic-transition spectra would change mainly
due to (1) the difference in the mean-square charge radii of thorium nuclei in the ground state and the isomeric state; and (2) the change in the hyperfine structure.

\begin{figure*}[!htpb]\centering
    \includegraphics[width=0.9\textwidth]{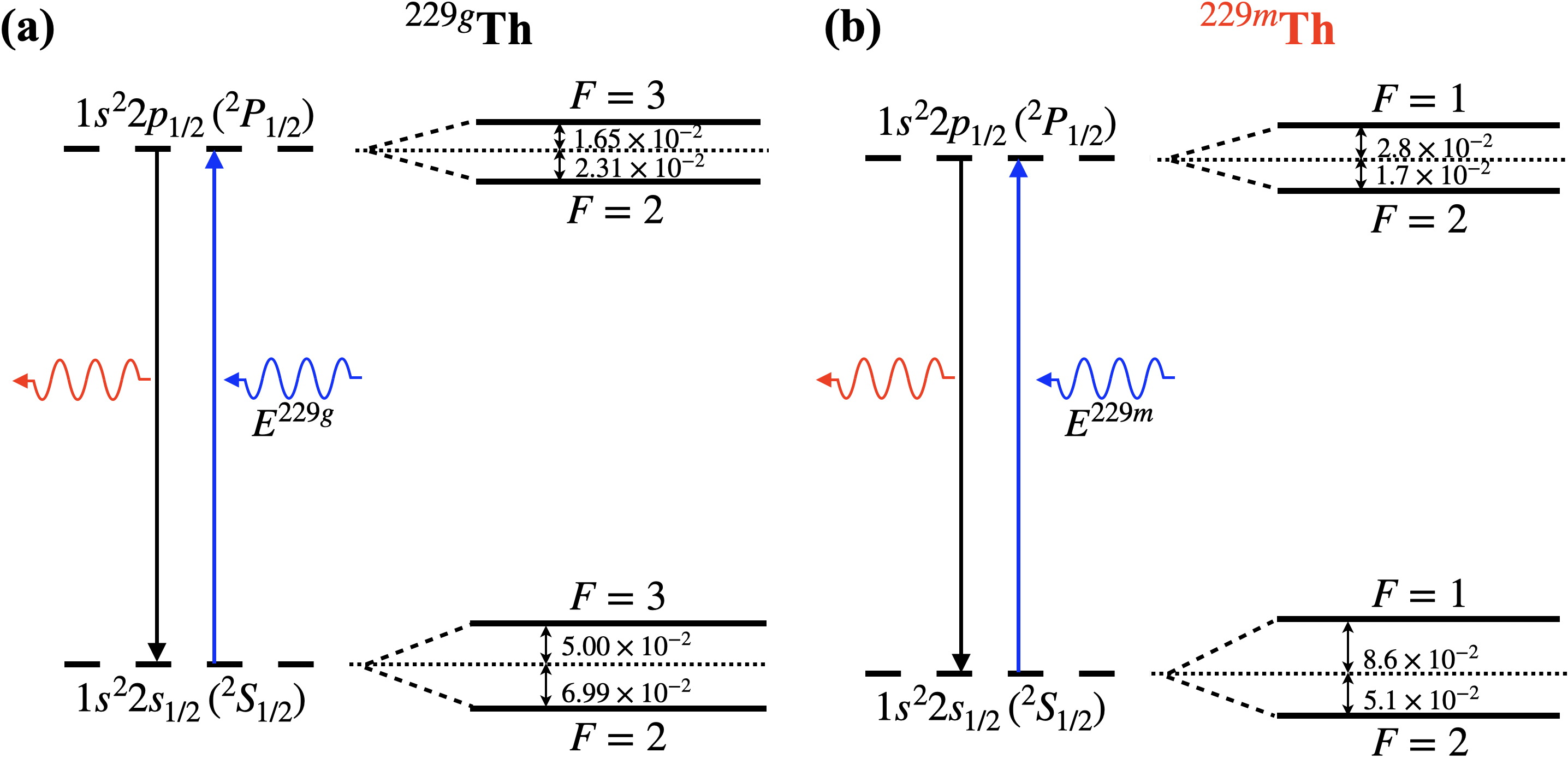}
    \caption{Detection of $^{229m}$Th via laser spectroscopy on the $^2S_{1/2}\rightarrow{}^2P_{1/2}$ electronic transition ($\approx271\,$eV) of Li-like thorium ions in
    (a) the nuclear ground state and (b) the isomeric state.
    The $^2S_{1/2}$ and $^2P_{1/2}$ energy levels without considering the HFS are denoted by dashed lines on the left.
    The HFS levels are shown as solid lines on the right and corresponding energy level splittings are indicated. All energies are given in eV.
    See the text for details of the transition energies, $E^{229g}$ and $E^{229m}$, and the hyperfine splittings.
    }
    \label{fig:Th_probe_elec}
\end{figure*}

We address in the following  Li-like $^{229}$Th ions and the electronic transition between the fine structure components $1s^22s_{1/2}$ ($^2S_{1/2}$) and $1s^22p_{1/2}$ ($^2P_{1/2}$); see Fig.\,\ref{fig:Th_probe_elec}.

\subsubsection{Isomer shift and hyperfine splitting}
\label{subsubsec:En_Li-like}
The energy of this transition in Li-like $^{232}$Th ions (in the nuclear ground state) has been calculated as 
$E^{232g} = 270.913(98)\,$eV
\cite{Yerokhin2018energy}.
The isotope shift of a transition can be obtained from
\begin{equation}
    \delta E^{A,A'}=2\pi\hbar\Delta \widetilde{K}^{RMS}\frac{M'-M}{M'M}+2\pi\hbar F_{FS} \delta\braket{r^2}^{A,A'}
\end{equation}
in the first-order perturbation-theory approximation \cite{Li2012IsotopeShift}.
Here, $\Delta \widetilde{K}^{RMS}$ is the relativistic mass-shift parameter, $M'$ and $M$ are the nuclear masses, $F_{FS}$ is the field-shift factor, and $\delta\braket{r^2}^{A,A'}=\braket{r^2}^{A}-\braket{r^2}^{A'}$ is the difference in the mean-square charge radii.
For Li-like Th$^{87+}$, we can use $\Delta \widetilde{K}^{RMS}=-7.832\times10^5$\,GHz$\cdot$u (u is the unified atomic mass unit) and $F_{FS}=-1.518\times10^5\,$GHz/fm$^2$ \cite{Li2012IsotopeShift}. The uncertainties of $\Delta \widetilde{K}^{RMS}$ and $F_{FS}$, which are not presented here, have negligible contribution to the error of the isomer shift $\delta E^{229m,229g}$ and the error of transition energies between hyperfine components of $^2S_{1/2}$ and $^2P_{1/2}$ in $^{229m,229g}$Th. The former is dominated by the uncertainty in $\delta\braket{r^2}^{229m,229g}$, while the latter is dominated by the uncertainty in $E^{232g}$.
Considering $\delta\braket{r^2}^{232g,229g}= 0.334(1)\,$fm$^2$ \cite{Angeli2013ChargeRadii} and  $\delta \braket{r^2}^{229m,229g} = 0.0105(13)\,$fm$^2$ \cite{Safronova2018ChargeRadii}, the energy of the transition $^2S_{1/2}\rightarrow {}^2P_{1/2}$ can be derived as $E^{229g}= 271.122(98)$\,eV and $E^{229m}=271.116(98)$\,eV for nuclei in the ground and isomeric state, respectively.
The isomeric shift is $\delta E^{229m,229g}=2\pi\hbar F_{FS} \delta\braket{r^2}^{229m,229g}=-0.0066(8)$\,eV.

Now we calculate the hyperfine splitting in Li-like $^{229}$Th ions. The $^2S_{1/2}$ and $^2P_{1/2}$ levels have two hyperfine components, $F=I-1/2,I+1/2$.
Since the electric quadrupole interaction is zero in the Li-like ion system with $J=1/2$, we only consider the magnetic dipole interaction between the electrons and the nucleus.
Therefore, the energy shifts of HFS have the form of $E_\textrm{hf}=A_\textrm{hf}\braket{\vec{I}\cdot\vec{J}}/\hbar^2=A_\textrm{hf}[F(F+1)-I(I+1)-J(J+1)]/2$. The $A_\textrm{hf}$ coefficients, as listed in Table\,\ref{tab:hfs_energy}, are calculated through the multi-configuration Dirac Fock (MCDF) method using the GRASP software package \cite{GRANT1980207}. The energy shifts of the $F=2$ levels due to the NHM are negligible, about $2\times10^{-4}$\,eV in Li-like $^{229}$Th ions \cite{Shabaev2021NHM}.

\begin{table}[htpb]
    \centering
    \begin{tabular*}{\linewidth}{@{\extracolsep{\fill}} lccc}
    \hline 
    \hline
    &HFS    &$A_\textrm{hf}\,(2\pi\hbar\times10^6$\,MHz)  &$E_\textrm{hf}$\,($10^{-2}\,$eV)\\
    \hline \\[-0.2cm]
    \multirow{4}{*}{$^{229g}$Th} &$^2S_{1/2}$, $F=2$     &9.66(19)   &-6.99(14) \\
    &$^2S_{1/2}$, $F=3$     &9.66(19)   &5.00(10) \\
    &$^2P_{1/2}$, $F=2$   &3.20(6)   &-2.31(4) \\
    &$^2P_{1/2}$, $F=3$   &3.20(6)   &1.65(3) \\
    \hline
    \\[-0.2cm]
    \multirow{4}{*}{$^{229m}$Th} &$^2S_{1/2}$, $F=1$     &-16.6(19)   &8.6(9) \\
    &$^2S_{1/2}$, $F=2$     &-16.6(19)   &-5.1(5) \\
    &$^2P_{1/2}$, $F=1$   &-5.5(8)    &2.8(4) \\
    &$^2P_{1/2}$, $F=2$   &-5.5(8)    &-1.7(2) \\
    \hline
    \hline
    \end{tabular*}
    \caption{The hyperfine splitting of $^2S_{1/2}$ and $^2P_{1/2}$ electronic levels of Li-like $^{229g(m)}$Th ions.
    }
    \label{tab:hfs_energy}
\end{table}

\subsubsection{Photoabsorption spectra with unpolarized and polarized nuclei}
\label{subsubsection:spectra_pol}

The excitation rate of 
the transition from a HFS level
$\ket{g}=\ket{^2S_{1/2},I,F_g}$
to
$\ket{e}=\ket{^2P_{1/2},I,F_e}$
is proportional to the
radiative width of the decay channel $\ket{e}\rightarrow\ket{g}$, given by \cite{Auzinsh2010LightAtom}
\begin{equation}
\label{Eq:nat_width}
    \Gamma_{F_e\rightarrow F_g}=g_{F_g} \cdot g_{J_e}
    \begin{Bmatrix}
    J_e & F_e &I\\
     F_g& J_g &L
    \end{Bmatrix}^2 \Gamma_{\xi_eJ_e\rightarrow \xi_gJ_g},
\end{equation}
where $g_{F_g}=2F_g+1$ and $g_{J_e} = 2J_e+1$ are the degeneracy factors,
$L=1$ is the multipolarity of the $E1$ transition,
$\Gamma_{\xi_eJ_e\rightarrow\xi_gJ_g}=1.12(1)\times10^{-5}\,$eV is the radiative width of the $^2P_{1/2}$ state \cite{Theodosiou1991Li-Ion}.
We list the $g_{F_e}\cdot\Gamma_{F_e\rightarrow F_g}$ values directly relevant to the relative excitation rates, and the transition energy $E_{if}$ in Table\,\ref{tab:hfs_tran} for $^{229g(m)}$Th ions.

\begin{table}[htpb]
    \centering
    \begin{tabular*}{\linewidth}{@{\extracolsep{\fill}} lccc}
    \hline 
    \hline
    &Transition    &$E_{if}$\,(eV)  &$g_{F_e}\cdot\Gamma_{F_e\rightarrow F_g}$\\
    &$^2S_{1/2}\rightarrow {}^2P_{1/2}$&  &($10^{-5}$\,eV)
    \\
    \hline \\[-0.2cm]
    \multirow{4}{*}{$^{229g}$Th} &$F_g=3\rightarrow F_e=2$    &271.049(98)   &4.34(4) \\
    &$F_g=3\rightarrow F_e=3$    &271.089(98)   &3.47(4) \\
    &$F_g=2\rightarrow F_e=2$    &271.169(98)   &1.24(1) \\
    &$F_g=2\rightarrow F_e=3$    &271.208(98)   &4.34(4) \\
    \hline \\[-0.2cm]
    \multirow{4}{*}{$^{229m}$Th} &$F_g=1\rightarrow F_e=2$    &271.012(98)   &2.79(2) \\
    &$F_g=1\rightarrow F_e=1$    &271.057(98)   &0.558(3) \\
    &$F_g=2\rightarrow F_e=2$    &271.149(98)   &2.79(2) \\
    &$F_g=2\rightarrow F_e=1$    &271.194(98)   &2.79(2) \\
    \hline
    \hline
    \end{tabular*}
    \caption{The energies ($E_{if}$) and $g_{F_e}\cdot\Gamma_{F_e\rightarrow F_g}$ of transitions between hyperfine components in Li-like $^{229g(m)}$Th ions.}
    \label{tab:hfs_tran}
\end{table}

\begin{figure}[!htpb]\centering
    \includegraphics[width=1.0\linewidth]{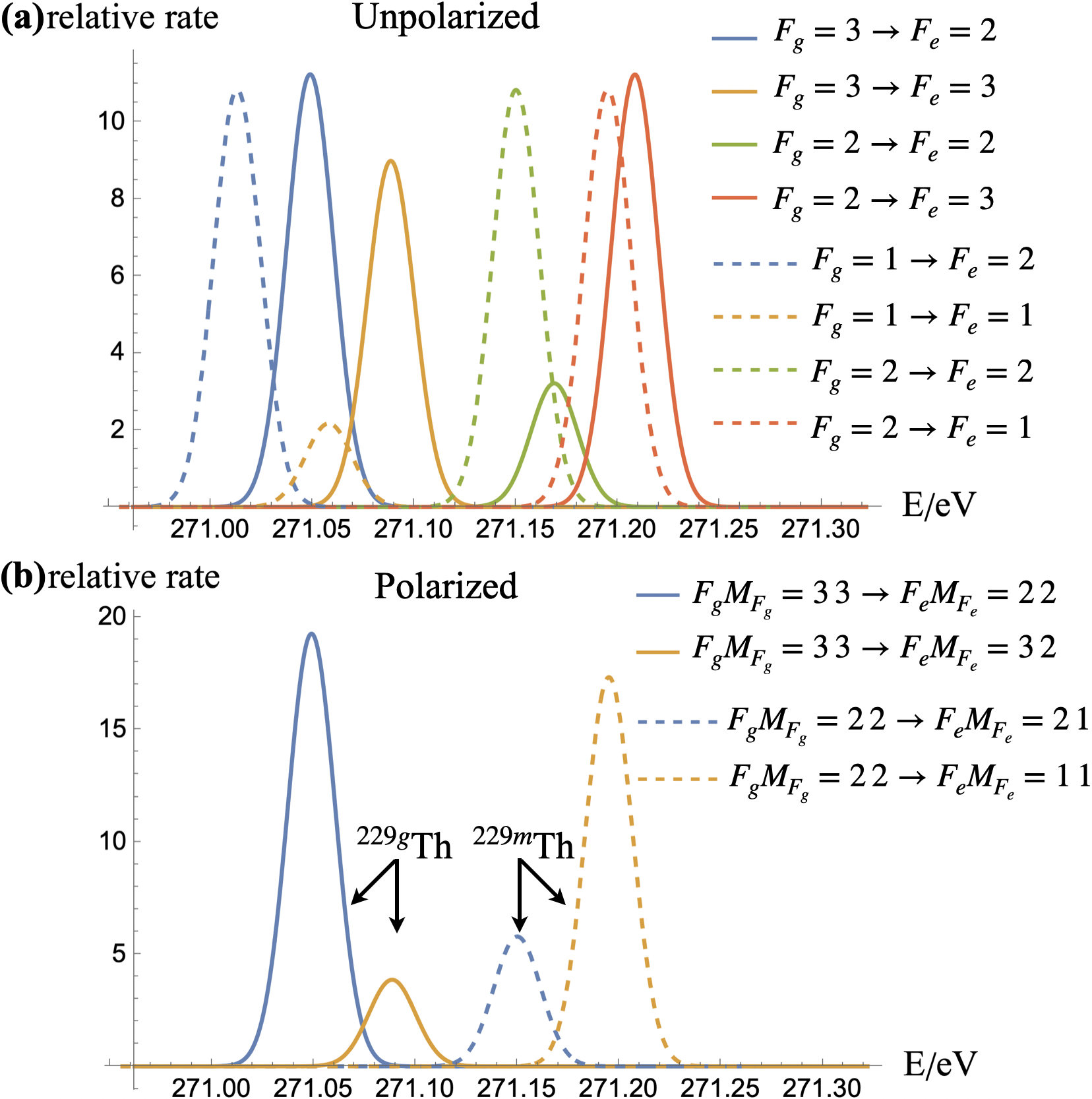}
    \caption{
    Relative excitation rates
    of (a) transitions from $\ket{^2S_{1/2},I,F_g}$ to $\ket{^2P_{1/2},I,F_e}$ in unpolarized Li-like $^{229}$Th ions; see Eq.\,\eqref{Eq:Exc_rate}, and (b) transitions from $\ket{^2S_{1/2},I, F_gM_{F_g}}$ to $\ket{{}^2P_{1/2},I, F_eM_{F_e}}$ in fully polarized Li-like $^{229}$Th ions; see Eq.\,\eqref{Eq:Exc_rate_pol}.
    The horizontal axis denotes photon energies in the ion frame. Solid lines and dashed lines are for ions with nuclei in the ground and isomeric state, respectively.
    }
    \label{fig:Probe_spec_Li-like}
\end{figure}

With the ion-energy spread (assumed to have a Gaussian line shape) taken into account, the relative excitation rate as a function of photon energies in the ion frame is
\begin{equation}
\label{Eq:Exc_rate}
    R_{\textrm{exc}}=
    \frac{g_{F_e}}{\sum\limits_{F_g} g_{F_g}}\frac{\Gamma_{F_e\rightarrow F_g}}{\Gamma_{\xi_eJ_e\rightarrow\xi_gJ_g}}\frac{1}{\sqrt{2\pi}\sigma_E}\exp\left\{-\frac{(E- E_{if})^2}{2\sigma_E^2}\right\}\,,
\end{equation}
where $2\sqrt{2\ln{2}}\sigma_E\approx(\Delta\gamma/\gamma)E_{if}$ is the FWHM energy of the transition resulting from the ion-energy spread, and the relative population of hyperfine level $F_g$ is assumed to be
proportional to the level degeneracy.
We note that the ion-frame lifetime of the HFS level $F_g=3$ in Li-like $^{229g}$Th is about 27\,s. The lifetimes of HFS levels $F_g=1$ and $2$ in Li-like $^{229m}$Th are about 2.2\,s and 0.88\,s, respectively, considering the NHM \cite{Shabaev2021NHM}.
The photon-absorption spectra in the ion frame for $^{229g}$Th and $^{229m}$Th are displayed in Fig.\,\ref{fig:Probe_spec_Li-like}\,(a) for $\Delta\gamma/\gamma=10^{-4}$.
Through tuning photon energies and monitoring the variation of detection rates of secondary photons, excitation of the isomeric state can be probed. For instance, with photon energies tuned to $271.01\cdot(1\pm0.5\Delta\gamma/\gamma)$\,eV in the ion frame, there is a significant contrast in excitation rates of ions in the nuclear ground state and the isomeric state. Essentially, only ions in the isomeric state can be excited (from $F_g=1$ to $F_e=2$).

Up to now, we have not discussed the effects of nuclear and electronic polarization.
In fact, it is possible to produce polarized PSI by optical pumping, see, for instance, Refs.\,\cite{Budker2020_AdP_GF,Bieron2022OpticalPump}.
They would be easier to probe using laser photons with suitable polarization.
Ions could be polarized in a single path through the interaction region with efficient optical pumping. If these ions are probed promptly, it may be unnecessary to preserve the polarization for a round trip in the storage ring.
Preserving the polarization of PSI in the ring is under investigation, see, for instance, Ref.\,\cite{Bondarevskaya2011_pol_theory}.

Suppose that we have produced fully polarized ions, where only the sublevel of $^2S_{1/2}$ with the maximum $M_{F_g}$ is populated. This could be realized via optical pumping using broadbrand $\sigma^+$ light to drive transitions between $^2S_{1/2}$ and $^2P_{1/2}$.
Then we can use $\sigma^-$ light for probing.
The transitions that will be excited in $^{229g}$Th are from $F_g,M_{F_g}=3,3$ to $F_e,M_{F_e}=2,2$ and $3,2$. For ions in the isomeric state, they are transitions from $F_g,M_{F_g}=2,2$ to $F_e,M_{F_e}=1,1$ and $2,1$.

Radiative widths of the corresponding decay channels from $F_e,M_{F_e}$ to $F_g,M_{F_g}$ are given by
\begin{align}\label{Eq:nat_width_sublevel}
    \Gamma_{F_eM_{F_e}\rightarrow F_gM_{F_g}}=
    &g_{F_e}\cdot g_{J_e}\begin{Bmatrix}
    J_e & F_e &I\\
    F_g & J_g &L
    \end{Bmatrix}^2 \Gamma_{\xi_eJ_e\rightarrow \xi_gJ_g}\nonumber\\
    & \cdot\left |\braket{F_eM_{F_e}L\lambda|F_gM_{F_g}}\right|^2
    \,,
\end{align}
where we have $L=1$ for $E1$ transitions and $\lambda=M_{F_g}-M_{F_e}$.
The relative excitation rate displayed in Fig.\,\ref{fig:Probe_spec_Li-like}\,(b) is
\begin{equation}
\label{Eq:Exc_rate_pol}
    R_{\textrm{exc}}= \frac{\Gamma_{F_eM_{F_e}\rightarrow F_gM_{F_g}}}{\Gamma_{\xi_eJ_e\rightarrow\xi_gJ_g}}\frac{1}{\sqrt{2\pi}\sigma_E}\exp\left\{-\frac{(E-E_{if})^2}{2\sigma_E^2}\right\}.
\end{equation}
We can see that the two spectra in the figure are well separated and therefore can be more easily distinguished than those in Fig.\,\ref{fig:Probe_spec_Li-like}\,(a).

Excitation of the 271\,eV transition can be implemented, for example, at the SPS or LHC. Generally, electronic transitions can be excited more efficiently compared to nuclear transitions. Significantly higher photon-detection rate can be obtained, thus affording a higher statistical precision in a given detection time.
The photon-detection rate using the electronic transition between $^2S_{1/2}$ and $^2P_{1/2}$
can reach $p_\textrm{d}\sim N_{\textrm{ion}}N_b f_b$, which is $p_\textrm{d}\sim 10^{14}\,$s$^{-1}$ and $10^{15}\,$s$^{-1}$ at the SPS and LHC, respectively.
Note that the nuclear isomeric state has a radiative lifetime of about $\gamma\cdot1\,$s in the lab frame. During the long lifetime, we can tune the laser to excite transitions between different HFS levels. Exciting different transitions, which could also be achieved using multiple lasers, is needed for continuous detection here because there is no cycling transition between HFS levels of ${}^2S_{1/2}$ and ${}^2P_{1/2}$.
Measuring the energy of the secondary photons, denoted as $E_{sps}$,
could determine $\gamma^\textrm{ave}$ with a high precision
according to $\gamma=(E_{sps}/4\hbar\omega_{sp})^{1/2}$,
where $\hbar\omega_{sp}$ is the lab-frame photon energy
of the laser exciting the electronic transition.
Depending on how the isomers are produced, the isomeric-state energy could be determined with a precision similar to that of $\gamma^\textrm{ave}$.

Another potential advantage of using PSI is that laser cooling via electronic transitions could be implemented
(see, for example, Ref.\,\cite{Krasny2020LaserCooling}), which would further reduce the ion-energy spread allowing for
better resolution of the transitions in Fig.\,\ref{fig:Probe_spec_Li-like}.
Laser cooling of ultrarelativistic Li-like Pb ions via the $^2S_{1/2}\rightarrow{}^2P_{1/2}$ electronic transition 
was proposed for SPS \cite{Krasny2019PoP}.
At lower energies, laser cooling of a bunched Li-like carbon and oxygen ion beam was already demonstrated at ESR and CSRe ion storage rings, see Refs.\,\cite{Wen2014_las_cool,wen2019laser} and the review \cite{Steck2020HeavyIonRing}.

\section{Search for the resonance}
\label{Sec:search_res}

In the direct excitation case, we need to scan across the possible energy of the isomeric state, which is currently known as $8.19(12)$\,eV \cite{Peik2021}, to search for the resonance. Besides tuning the energy of laser photons, we can also tune the energy of relativistic HCI, i.e., the relativistic Lorentz factor $\gamma$, and the ion energy spread.
We can start with ions having a relative energy spread of  $\Delta\gamma/\gamma\approx10^{-3}$, and tune $\gamma$ to scan across the energy range, $8.19\pm(3\times0.12)$\,eV, which would require about 90 scan steps.
The resonant excitation rate of isomers could be sufficiently high for experiments even for bare nuclei, where there is no NHM and the isomeric state has a narrow radiative width.
For instance, it takes around 0.6\,s to excite a significant fraction of ions in an ion bunch composed of bare nuclei at the SPS for $\Delta\gamma/\gamma=10^{-3}$ (see also the examples in Sec.\,\ref{subsubsec:reso_multi_pulse}).
Detection of isomer excitation can be done by using the schemes presented in Sec.\,\ref{Sec:Detect}.

After finding the resonance, i.e, finding the $\gamma$ required for $E_{\textrm{iso}}=2\gamma\hbar\omega$, the transition energy is known with a relative precision of about $\Delta\gamma/\gamma\approx10^{-3}$.
At the second stage, we can reduce the relative ion-energy spread from $10^{-3}$ to $10^{-4}$, and, again, tune the relativistic factor $\gamma$ to scan across the narrower energy range.
Through this process, the transition energy would be determined with a precision improved by one order of magnitude.
One can continue such procedure if the ion-energy spread can
be further reduced.

Determining the ion-frame transition energy requires knowing the central value of $\gamma$, $\gamma^{\textrm{ave}}$, with a high precision, which could be done via, for example, measuring secondary photons (see Ref.\,\cite{Budker2020_AdP_GF} and schemes in Sec.\,\ref{Sec:Detect}).
Prior to precision measurements of $\gamma^{\textrm{ave}}$,
we should achieve a better match between the photon-absorption spectrum considering the ion-energy spread and the laser-pulse spectrum in the ion frame in terms of their centers, i.e.,
$|E_\textrm{iso}-2\gamma^{\textrm{ave}}\hbar\omega|/E_\textrm{iso}\ll\Delta\gamma/\gamma$.
This could be accomplished by utilizing a laser with a narrow bandwidth and scanning the laser-photon energy $\hbar\omega$, maximizing the isomer-production rate.
Matching the two spectra would allow more precise determination of the isomeric-state energy via measuring $\gamma^{\textrm{ave}}$; see Sec.\,\ref{Sec:Detect}.

As mentioned above, further improvement
could be granted by employing laser cooling
of PSI, for instance, as projected for Li-like lead ions at the SPS \cite{Krasny2019PoP}, which could reduce the ion-energy spread by another one or two orders of magnitude.

The transitions related to an intermediate electronic excited state discussed in Sec.\,\ref{Subsec:exc_271eV}
have a relative energy uncertainty of about $5\times10^{-4}$.
When exciting these transitions, one can start with ions having a relative energy spread of $\Delta\gamma/\gamma\approx10^{-4}$ to search for the resonance.
For transitions discussed in Sec.\,\ref{Subsec:exc_29keV}, the relative uncertainty of the transition energy between the nuclear ground (isomeric) state and the 29\,keV state is about $2.4\times10^{-6}$ ($4.8\times10^{-6}$), which is more precise than the typical ion energy spread, $\Delta\gamma/\gamma\approx10^{-3}-10^{-5}$ in storage rings. Therefore, these transitions could be excited directly without varying $\gamma$ to scan across a large energy range.
In these cases, we also need to match the photon-absorption spectrum and the laser-pulse spectrum, to drive the transition efficiently and to facilitate further measurement of the isomeric-state energy.

\section{other low-lying nuclear transitions}
\label{Sec:other_low_state}

The excitation and probing schemes of $^{229m}$Th discussed above could also be applied to other low-lying nuclear states with experimental parameters regarding the laser beam and ion bunch chosen case by case.
$^{229}$Th is unique in featuring a nuclear excited state in the eV range. The next confirmed low-energy nuclear excitation occurs in $^{235}$U at 76\,eV \cite{NNDC}.
The extremely narrow radiative width ($\Gamma_{\textrm{rad}}/\hbar \approx 9.9\times10^{-25}\,$s$^{-1}$) and $E3$ multipolarity \cite{Budker2021GF_nucl} discourage laser excitation of the isomeric state.
A survey of other relatively low-lying nuclear transitions can be found in Ref.\,\cite{Budker2021GF_nucl}, where the proposed Gamma Factory can access those transition energies.

Notably, there exists another interesting case, $^{229}$Pa (a nucleus with the same mass number $A$ as $^{229}$Th but one more proton).
$^{229}$Pa is expected to have octupole deformation in its nuclear ground state ($I^P=5/2^+$), which may have a nearly-degenerate first excited state of opposite parity ($I^P=5/2^-$), with an excitation energy as low as $60\pm50$\,eV \cite{Ahmad1982Pa,Ahmad2015Pa}.
This would enhance the nuclear Schiff moment \cite{Flambaum2020_Schiff_moment} and make $^{229}$Pa orders of magnitude more sensitive to hadronic CP violation compared to other nuclei such as $^{199}$Hg \cite{Singh2019T-vio,Chishti2020_edm}.
However, the existence of the low-lying nuclear excited state has not been unambiguously proven. The large error in a recent analysis ($60\pm50$\,eV \cite{Ahmad2015Pa}) still makes the existence of the ground-state doublet uncertain.

Laser spectroscopy of relativistic highly charged $^{229}$Pa ions could provide alternative approaches to probing the doublet \cite{Singh2021}. The $I^P=5/2^-$ state could be excited by either direct resonance excitation or excitation via an intermediate state.
An $E1$ transition could be allowed between the nuclear ground and first excited states.

For direct laser excitation, we need to scan across a large energy range, $60\pm50\,$eV, to search for the resonance. This can be realized, for example, at SPS by tuning the relativistic factor of the HCI from $\gamma=10$ to 110 together with a laser having a fixed energy of about 0.5\,eV in the lab frame. It could take up to about 2400 scan steps assuming using ions with a relative energy spread of $\Delta\gamma/\gamma=10^{-3}$.
One could probe the existence of the doublet and the energy splitting via measuring the radiative decay of the $I^P=5/2^-$ state.
Furthermore, measurement of an enhanced $E3$ transition strength between the doublet would indicate the octupole collectivity of $^{229}$Pa unambiguously.

Besides, one could also excite the transition from the nuclear ground state to a higher nuclear excited state at 11.4(2) or 11.6(3)\,keV \cite{NNDC}. Measuring radiative decay of the higher state could determine whether there is a decay channel into a low-lying nuclear state and probe the energy splitting of the ground-state doublet.
The expected small energy splitting in the ground state doublet, if it exists, would enhance the NHM effect in H- or Li-like $^{229}$Pa ions, which might offer more possibilities for laser excitation of the first excited state, similar to those discussed in Sec.\,\ref{subsubsec:res_H-like} and \ref{Subsec:exc_271eV}.
We can also detect the change of the nuclear state via laser spectroscopy on electronic transitions, following the discussion in Sec.\,\ref{Subsec:elec_tran}.

We note that neutral $^{229g}$Pa atoms in the nuclear ground state have a half-life of about 1.5(5)\,d\,\cite{ENSDF} due to electron capture and $\alpha$-decay which would pose a severe challenge for producing a beam from an ion source. Hence, $^{229}$Pa ions need to be freshly produced at a radioactive ion beam facility. Projectile fragmentation of $^{238}$U is well suited for this purpose\,\cite{LISE}. Assuming the same beam/target parameters as in Sec.\,\ref{Sec:ion_source}, we obtain $6\cdot10^6$ $^{229}$Pa ions produced per pulse. Large amounts of $^{229}$Pa can be produced at FRIB\,\cite{Abel2019_PaSource} and RIKEN\,\cite{Kubo-2003} where the worldwide highest uranium primary beam intensities are available. The advantage of using GSI/FAIR, HIAF or CERN facilities would be the higher projectile energy and thus the ability to efficiently produce fully-stripped ions. Furthermore, storage facilities would be available as discussed above for the $^{229}$Th ions. In bare nuclei the 99.52(5)\% electron capture branch\,\cite{ENSDF} is disabled correspondingly increasing the half-life of $^{229g}$Pa$^{91+}$ nuclei. Moreover, if stored at high energy in a storage ring, the half-life is further increased in the lab frame due to the relativistic time dilation.

\section{Conclusion}

We propose multiple approaches towards laser excitation of the nuclear isomeric state in relativistic highly charged $^{229}$Th ions at high-energy storage rings. It can be achieved
via direct resonant excitation, excitation via an intermediate electronic excited state, or excitation via an intermediate nuclear excited state.
Due to the Lorentz boost of laser photon energies in the ion frame, lasers with wavelengths in the visible range or longer can be used for excitation.
Searching for the resonance and matching the laser-pulse spectrum and the photon-absorption spectrum in the ion frame can be realized through scanning the ion energy, adjusting the ion-energy spread as well as tuning the laser beam.

Among the discussed excitation schemes, STIRAP excitation via the 29\,keV state may require further development of the laser or storage-ring technologies. Schemes utilizing repeated pumping relax the demand on the laser-pulse intensity.
Nuclear-hyperfine-mixing assisted excitation schemes are also discussed, and could be most favorable for experimental implementation.
In the direct resonant excitation scenario, the significantly reduced isomeric-state lifetime in H- or Li-like $^{229}$Th ions corresponds to much higher excitation rates compared to bare thorium nuclei, thus making it more practical to implement a $\pi$ pulse.
Furthermore, the NHM effect induces $E1$ transitions that change both the electronic and nuclear states, enabling efficient nuclear isomer excitation via an intermediate electronic excited state in Li-like $^{229}$Th. The transition energies are also easily accessible by x-ray lasers, possibly allowing for isomer excitation in static $^{229}$Th ions using x-ray pulses.

We also present schemes to probe the produced $^{229m}$Th isomers, exploiting the radiative decay of the isomeric state,
radiative decay of the second nuclear excited state,
or laser spectroscopy of electronic transitions.
Among those detection schemes, the NHM effect helps increase the photon detection rate of the isomer decay,
the precise knowledge of the second nuclear excited state energy can be exploited to measure the isomeric-state energy precisely, and using electronic transitions leads to higher photon detection rates than nuclear transitions.
Our estimates show that the isomeric-state energy can be determined with orders-of-magnitude improvement in precision compared to the current result.

The schemes proposed here for $^{229}$Th could also be adapted to low-energy nuclear states in other nuclei including $^{229}$Pa.

\section*{Acknowledgements}

The authors are grateful to Mei Bai, Carsten Brandau, Wick Haxton,
M. Witold Krasny, Anton Peshkov, Alexey Petrenko,
Jaideep Singh, Markus Steck, and Thomas St{\"o}hlker for helpful discussions. This work was supported in part by the Deutsche Forschungsgemeinschaft (DFG, German Research Foundation) Project ID 390831469:  EXC 2118 (PRISMA+ Cluster of Excellence) and the Helmholtz
Excellence Network ExNet02. A.~P\'alffy gratefully acknowledges funding from the DFG in the framework of the Heisenberg Program.
AS acknowledges support by the DFG under Germany's Excellence Strategy EXC-2123 QuantumFrontiers-390837967.
Yury Litvinov acknowledges support by the State of Hesse within the Research Cluster ELEMENTS (Project ID 500/10.006).
PGT and A. P\'alffy acknowledge support via the ‘ThoriumNuclearClock’ project that has received funding from the ERC under the European Union’s Horizon 2020 research and innovation programme (Grant Agreement No. 856415).

\bibliography{GF_Th229}
\end{document}